# Defect chemistry of mixed ionic-electronic conductors under light: halide perovskites as master example

Davide Moia,* Joachim Maier

Max Planck Institute for Solid State Research, Heisenbergstraße 1, 70569 Stuttgart, Germany

*moia.davide@gmail.com,
Current address: Fluxim AG, Katharina-Sulzer-Platz 2, 8400 Winterthur, Switzerland

**Abstract**

Shining light on a mixed ionic-electronic conductor induces variations in both its electronic and ionic behaviors. While optoelectronic processes in semiconductors with negligible ionic conductivities are well understood, the role of mobile ions in photo-active mixed conductors, such as hybrid halide perovskites, is largely unexplored. Here, we propose a model addressing this problem, relating optoelectronics and optoionics. Using methylammonium lead iodide (MAPI) as model material, we discuss the expected influence of optical bias on the material's charge carrier chemistry under steady-state conditions. We show that changes in the concentration of ionic defects under light with respect to the dark case are a direct consequence of their coupling to electrons and holes through the component chemical potential (here iodine) and the electroneutrality condition. Based on the trend in the quasi-Fermi level splitting in MAPI, we emphasize implications of controlling point defect chemistry for the function and performance optimization of solar energy conversion devices based on halide perovskites. Lastly, we show that in the presence of multiple redox reactions mediating the iodine quasi-equilibrium, either positive or negative changes in the ionic defect pair chemical potential can be obtained. These findings indicate the intriguing possibility to increase or to reduce ionic defect concentrations in mixed conductors through exposure to light.

**Introduction**

The efficacy of photoelectrochemical devices for energy applications based on semiconducting and mixed ionic-electronic conducting materials relies on controlling the behavior of and interactions between ions and electrons to convert, transfer and store energy. Understanding these aspects is crucial to the optimization of devices ranging from solar cells, batteries and fuel cells. The description of these systems can be greatly simplified, if the condition of local equilibrium applies during their operation. It is then straightforward to define the thermodynamic state at each position in the device through parameters including the (electro)chemical potential for all charged and neutral species, stoichiometry, defect formation energies to name a few. The quantification of such parameters is still possible locally, when a chemical or electrochemical bias applied across the system induces gradients in the electrochemical potentials of electrons and ions (e.g. batteries or fuel cells under operation).[1]

In many cases of interest, the applied bias leads to deviation from local equilibrium. This is the case for semiconductors used in solar cells, where light absorption leads to local non-equilibrium between the electronic charges populating different energy bands.[2,3] Local non-equilibrium can also arise in the dark, when a voltage bias is applied to such devices, due to injection selectivity of the contacts. A similar condition may, in principle, be obtained for different ionic defects, if the relevant sublattices are not at equilibrium with each other. The discussion of these situations can be addressed using quasi-equilibrium arguments, where different defects related to a specific component (e.g. conduction band electrons, $e'$, and valence band holes, $h^·$, associated with electrons $e^-$) are described by separate occupation statistical functions and separate quasi-electrochemical potentials. This treatment is applicable if equilibration within each charge carrier population occurs at a much faster rate than any of the reactions in which such carriers are involved.

In general, given a material component or defect $j$, its quasi-electrochemical potential can be expressed in the form:

$$\tilde{\mu}_j^* = \tilde{\mu}_{j,\text{eq}} + \delta\tilde{\mu}_j, \quad (1)$$

where the equilibrium value and the deviation from it are indicated with $\tilde{\mu}_{j,\text{eq}}$ and $\delta\tilde{\mu}_j$, respectively.

While these are not simple energies, the quasi-electrochemical potentials of electronic charges are often discussed in terms of quasi-Fermi energies for electrons, $E_{\text{F}n}$, and for holes, $E_{\text{F}p}$, each referring to separate Fermi-Dirac statistics.[4,5] The resulting quasi-Fermi level splitting $QFLS = E_{\text{F}n} - E_{\text{F}p}$ can then be expressed as the combined electron-hole chemical potential change according to

$$QFLS = \delta\tilde{\mu}_{e'} + \delta\tilde{\mu}_{h^{\cdot}}, \quad (2)$$

and it can be used to quantify the "degree of local electronic non-equilibrium". Assuming a dilute situation, the product of the concentrations of electrons ($n$) and holes ($p$) is related to the $QFLS$, as described by the modified mass-action law,

$$np = K_B \exp\left(\frac{QFLS}{k_B T}\right) \quad (3)$$

where $K_B$ is the mass action constant of $e'$-$h^{\cdot}$ thermal generation and recombination, often written as $n_i^2$, $k_B$ and $T$ are Boltzmann's constant and temperature. At equilibrium ($n = n_{\text{eq}}, p = p_{\text{eq}}$), this expression reduces to the conventional mass-action law $n_{\text{eq}} p_{\text{eq}} = K_B$ ($QFLS = 0$). For situations where all ionic defects are immobile, as is the case for most semiconductors used in solar cells, Equation 3 reflects the local quasi-equilibrium in the material under bias.

In a mixed ionic-electronic conductor, where electronic but also ionic charges are mobile, the equilibrium situation is defined by a more complex set of equations. These are derived from the mass-action laws involving ionic and electronic defects, coupled through the chemical potential (partial pressure) of components associated with the relevant mobile ions.[6–8] Under bias, the coupling between electrons and ions in these materials means that the ionic situation can vary too, even when a purely optoelectronic excitation is considered.

Hybrid halide perovskites are a relevant example to this question. These materials are used as active layers of high performance optoelectronic devices while also showing significant ionic conduction even at room temperature.[9,10] In methylammonium lead iodide (MAPbI$_3$, or MAPI), a reference compound for hybrid perovskite photovoltaics, migration of iodide defects (specifically vacancies) enables access to its defect chemical behavior by varying the iodine partial pressure in the system ($P(I_2)$).[3,11,12] While the characterization of MAPI under equilibrium conditions is becoming established, the study of its properties under light has resulted in numerous peculiar observations including anomalies in mass transport, phase stabilities and mechanical properties. Many of them have been interpreted in terms of coupled ionic-electronic effects and are still matter of debate.[13–18]

Other "unusual" photo-electrochemical effects in various mixed conductors have also been interpreted based on interactions between photo-generated electronic carriers and ionic defects (photo-ionic or optoionic effects).[14,19–22] The study of light effects in strontium titanate (STO)[19] highlighted the enhanced kinetics of oxygen incorporation in the material by (above bandgap) illumination, explained by the electronic contribution to the exchange reaction. While these investigations showed the effect of light on the surface kinetics, later reports discussed stoichiometric changes in conductivity on illumination for MAPI or STO.[14,22–24]. Light effects have also been reported as far as the grain boundary resistance (for Gd-doped ceria) is concerned.[21]

All these reports point towards fundamental interactions between electronic and ionic charges, which, on drastic increase of electronic charge carrier concentration under light, inevitably modify the ionic equilibrium too. While these studies have suggested models that could explain the experimental results, systematically treating the quasi-equilibrium behavior of mixed conductors under light bias means entering largely unknown territory.

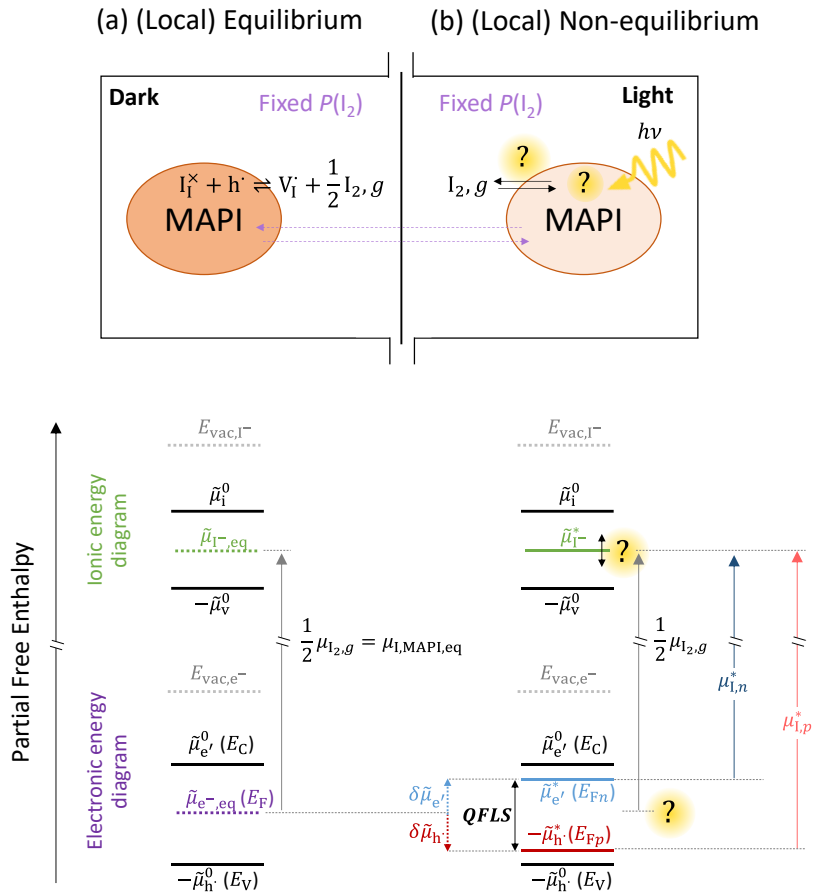

Figure 1. (top) Schematics of a mixed ionic-electronic conductor (in this case MAPI) and (bottom) generalized energy diagram including the electronic and ionic (iodide) defect energy levels (a) under equilibrium and (b) under illumination (out-of-equilibrium). In the top schematics, the two MAPI samples are at equilibrium with each other only as far as the exchange of iodine is concerned (see dashed purple arrows). Specifically, they do not exchange radiation with each other. The $P(I_2)$ dependence of the defect concentration under equilibrium can be evaluated based on mass-action laws and the exchange reaction with iodine in the gas phase. Such dependence is more complex for the quasi-equilibrium situation under light. In other words, while in (a) the relation between the electronic and the ionic electrochemical potentials and the chemical potential of iodine is straightforward ($\frac{1}{2}\mu_{I_2,g} = \tilde{\mu}_{I^-,eq} - \tilde{\mu}_{e^-,eq}$), in (b) such relation is not clearly defined, even assuming a single value for $\tilde{\mu}_{I^-}^*$ (quasi-chemical potentials of iodine defined as $\mu_{I,p}^* = -\tilde{\mu}_{V_I}^* + \tilde{\mu}_{h^\cdot}^*$ and $\mu_{I,n}^* = -\tilde{\mu}_{V_I}^* - \tilde{\mu}_{e'}^*$, where, $\tilde{\mu}_{I^-}^* = -\tilde{\mu}_{V_I}^*$).

Figures 1 display the research question of this study (top schematics): Given a mixed conductor (e.g. MAPI) and the point defect model describing its equilibrium situation (*e.g.* $P(I_2)$ dependence of defect concentrations, cf. dashed lines in Figure 3), how to describe the situation under light? This question is addressed at the bottom of Figure 1 based on the generalized energy level representation, which considers the standard partial (free) energy levels and electrochemical potentials associated with the electronic and the ionic (here iodide only, for simplicity) defects.[7] In the diagram, the position of the electrochemical potentials $\tilde{\mu}_{I^-}$ and $\tilde{\mu}_{e^-}$ relative to the standard potentials ($\tilde{\mu}^0$) determines the concentrations of iodide vacancies and interstitials, $V_I^{\cdot}$ and $I_i'$, and of electrons and holes, $e'$ and $h^{\cdot}$, respectively.

The chemical potential of iodine in MAPI corresponds to $\mu_{I,MAPI} = \tilde{\mu}_{I^-} - \tilde{\mu}_{e^-}$ and, at equilibrium, it is equal to the given chemical potential of iodine in the gas phase, $\mu_{I,MAPI,eq} = \frac{1}{2}\mu_{I_2,g}$. Taking the former relation for granted also for the nonequilibrium situation implies that changes in the electrochemical potential of electronic charges due to an applied bias result in changes in the iodine and/or iodide (electro)chemical potential in the material. The situation is complicated by the fact that electrons and holes are now not in equilibrium, as indicated by the two quasi-electrochemical potentials $\tilde{\mu}_{e'}^*$ and $\tilde{\mu}_{h^{\cdot}}^*$. Furthermore, ionic defect concentrations may also vary under bias. Therefore, the straightforward definition of $\mu_{I,MAPI}$ used above is no longer applicable (Figure 1b).

One way to describe this situation involves the definition of iodine quasi-chemical potentials, such as $\mu_{I,p}^* = -\tilde{\mu}_{V_I^{\cdot}}^* + \tilde{\mu}_{h^{\cdot}}^*$ and $\mu_{I,n}^* = -\tilde{\mu}_{V_I^{\cdot}}^* - \tilde{\mu}_{e'}^*$ (assuming $-\tilde{\mu}_{V_I^{\cdot}}^* = \tilde{\mu}_{I_i'}^* = \tilde{\mu}_{I^-}^*$), which would be identical at equilibrium but different otherwise (see also discussion by Kim *et al*.[14] and by Viernstein *et al*.[23] for halide and oxide perovskites, respectively). Their relation with the chemical potential of iodine in the gas phase is then kinetically determined. Further complexity is introduced by the influence of mobile ionic defects on the optoelectronic quasi-equilibrium (*e.g.* the electron-hole recombination rate).[25–29] A model to evaluate such interplays and to predict the electronic and ionic charge carrier concentrations as a function of the components partial pressures (*e.g.* $P(I_2)$ in MAPI) in a mixed conductor is currently missing.[30]

Here, we address the problem shown in Figure 1 by coupling the defect chemical relations describing ionic and electronic disorder in a mixed ionic-electronic conductor and the component quasi-equilibrium with the gas phase, with the equations associated with the generation and recombination of electronic charge carriers. We present results obtained using MAPI as model system and illustrate the expected trend in defect concentrations under illumination. In the framework of the quasi-equilibrium considerations, we discuss implications of these results for the study of photo-active mixed conductors and their use in photoelectrochemical devices for energy conversion. By exploring the effect of relevant kinetic and thermodynamic parameters on the material's defect chemistry under light, special emphasis is given to the predicted possibility to control (increase but also decrease) ionic defects concentration using light.

**The model**

The model presented in this study accounts for internal defect reactions occurring in a mixed ionic-electronic conductor exposed to light and for the exchange quasi-equilibria with the gas phase. We refer to a pore-free thin film of MAPI exposed to a set illumination condition and to a fixed iodine partial pressure, $P(I_2)$, at room temperature. We assume the film to be thin enough, so that, when exposed to light, the photo-generation rate can be assumed constant throughout the film volume. We also focus on a reaction-limited model, where transport is fast enough, resulting in homogeneous concentration profiles in the film for all mobile defects. Because of the complexity of including all possible defects in MAPI and the reactions involving them, we limit our analysis to the study of iodine and iodide defects, interstitial and vacancies. Regarding the solid-gas exchange of iodine, we focus on processes that involve neutral iodine defects only ($I_i^\times, V_I^\times$), while other reactions may also contribute to the real system.

Table 1. Defect chemical reactions used in this study to describe the electronic and ionic disorder in MAPI under constant $P(I_2)$ at equilibrium in the dark or under light. The expressions for the forward and backward rates are shown, as well as the equilibrium (eq) mass-action laws and the pseudo mass-action laws for the nonequilibrium case. The terms in () are used as labels to the defect reactions (cf. Figure 2; B = bandgap excitation, $\bar{F}$ = anti-Frenkel ionic defects, $\bar{F}^\times$ = neutral anti-Frenkel ionic defects, sg = solid-gas exchange, $n$ = electrons, $p$ = holes, v = vacancies, i = interstitials). [] indicates defect concentrations. $E_g$, $\Delta G_{\bar{F}}^0$ and $\Delta G_{\bar{F}^\times}^0$ are the standard free enthalpy of reactions $(B)$, $(\bar{F})$ and $(\bar{F}^\times)$.

| Reaction | Defect reaction | Rates | (Pseudo) mass-action law |
|---|---|---|---|
| Electronic | $h\nu \rightleftarrows e' + h^\cdot$  (B)<br>(radiative case) | $G = G_{ext} + \sum_k G_{th,k}$<br>$R = \sum_k R_k$<br>$k = rad, SRH, Aug, I$ | $K_B = n_{eq} p_{eq} \propto \exp\left(-\frac{E_g}{k_B T}\right)$<br>(light: $np = K_B \exp\left(\frac{QFLS}{k_B T}\right)$ ) |
| Ionic<br>(anti-Frenkel disorder) | $V_i^\times + I_I^\times \rightleftarrows I_i' + V_I^\cdot$  $(\bar{F})$ | $\vec{R}_{\bar{F}} = \vec{k}_{\bar{F}}$<br>$\overleftarrow{R}_{\bar{F}} = \overleftarrow{k}_{\bar{F}}[I_i'][V_I^\cdot]$ | $K_{\bar{F}} = [I_i']_{eq}[V_I^\cdot]_{eq} \propto \exp\left(-\frac{\Delta G_{\bar{F}}^0}{k_B T}\right)$<br>(light: $[I_i'][V_I^\cdot] = K_{\bar{F}} \exp\left(\frac{\Delta \mu_{\bar{F}}}{k_B T}\right)$ ) |
| Ionic<br>(neutral defects disorder) | $V_i^\times + I_I^\times \rightleftarrows I_i^\times + V_I^\times$  $(\bar{F}^\times)$ | $\vec{R}_{\bar{F}^\times} = \vec{k}_{\bar{F}^\times}$<br>$\overleftarrow{R}_{\bar{F}^\times} = \overleftarrow{k}_{\bar{F}^\times}[I_i'][V_I^\cdot]$ | $K_{\bar{F}^\times} = [I_i^\times]_{eq}[V_I^\times]_{eq} \propto \exp\left(-\frac{\Delta G_{\bar{F}^\times}^0}{k_B T}\right)$<br>(light: $[I_i^\times][V_I^\times] = K_{\bar{F}^\times} \exp\left[\frac{\Delta \mu_{\bar{F}^\times}}{k_B T}\right]$ ) |
| Redox<br>(interstitial) | $I_i' + h^\cdot \rightleftarrows I_i^\times$  $(p, i)$ | $\vec{R}_{p,i} = \vec{k}_{p,i}[I_i']p$<br>$\overleftarrow{R}_{p,i} = \overleftarrow{k}_{p,i}[I_i^\times]$ | $K_{p,i} = \frac{[I_i^\times]_{eq}}{[I_i']_{eq} p_{eq}}$ |
|  | $I_i' \rightleftarrows I_i^\times + e'$  $(n, i)$ | $\vec{R}_{n,i} = \vec{k}_{n,i}[I_i']$<br>$\overleftarrow{R}_{n,i} = \overleftarrow{k}_{n,i}[I_i^\times]n$ | $K_{n,i} = \frac{[I_i^\times]_{eq} n_{eq}}{[I_i']_{eq}} = K_{p,i} K_B$ |
| Redox<br>(vacancy) | $V_I^\times + h^\cdot \rightleftarrows V_I^\cdot$  $(p, v)$ | $\vec{R}_{p,v} = \vec{k}_{p,v}[V_I^\times]p$<br>$\overleftarrow{R}_{p,v} = \overleftarrow{k}_{p,v}[V_I^\cdot]$ | $K_{p,v} = \frac{[V_I^\cdot]_{eq}}{[V_I^\times]_{eq} p_{eq}}$ |
|  | $V_I^\times \rightleftarrows V_I^\cdot + e'$  $(n, v)$ | $\vec{R}_{n,v} = \vec{k}_{n,v}[V_I^\times]$<br>$\overleftarrow{R}_{n,v} = \overleftarrow{k}_{n,v}[V_I^\cdot]n$ | $K_{n,v} = \frac{[V_I^\cdot]_{eq} n_{eq}}{[V_I^\times]_{eq}} = K_{p,v} K_B$ |
| Iodine exchange<br>(interstitial-mediated) | $I_i^\times \rightleftarrows V_i^\times + \frac{1}{2} I_2, g$  (sg, i) | $\vec{R}_{sg,i} = \vec{k}_{sg,i}[I_i^\times]$<br>$\overleftarrow{R}_{sg,i} = \overleftarrow{k}_{sg,i} P(I_2)^{\frac{1}{2}}$ | $K_{sg,i} = \frac{P(I_2)^{\frac{1}{2}}}{[I_i^\times]_{eq}}$ |
| Iodine exchange<br>(vacancy-mediated) | $I_I^\times \rightleftarrows V_I^\times + \frac{1}{2} I_2, g$  (sg, v) | $\vec{R}_{sg,v} = \vec{k}_{sg,v}$<br>$\overleftarrow{R}_{sg,v} = \overleftarrow{k}_{sg,v} P(I_2)^{\frac{1}{2}}[V_I^\times]$ | $K_{sg,v} = P(I_2)^{\frac{1}{2}}[V_I^\times]_{eq} = K_{sg,i} K_{\bar{F}^\times}$ |

The relevant reactions are described in Table 1 and schematically represented in Figure 2. For reaction $r$, the expressions for the rate of forward and backward reactions used in the kinetic model, $\vec{R}_r$ and $\overleftarrow{R}_r$, are defined based on rate constants ($\vec{k}_r$ and $\overleftarrow{k}_r$) and the concentration of the relevant reactants. We indicate the mass-action constant $K_r$, defined based on the equilibrium concentrations of the relevant defects ($K_r = \vec{k}_r/\overleftarrow{k}_r$).[23] On a first order approximation, and based on the quasi-equilibrium framework, we can assume that the value of the rate constants for the forward and the backward reactions are the same under bias and under dark conditions. All relevant input parameters are shown in section 1 of the Supporting Information.

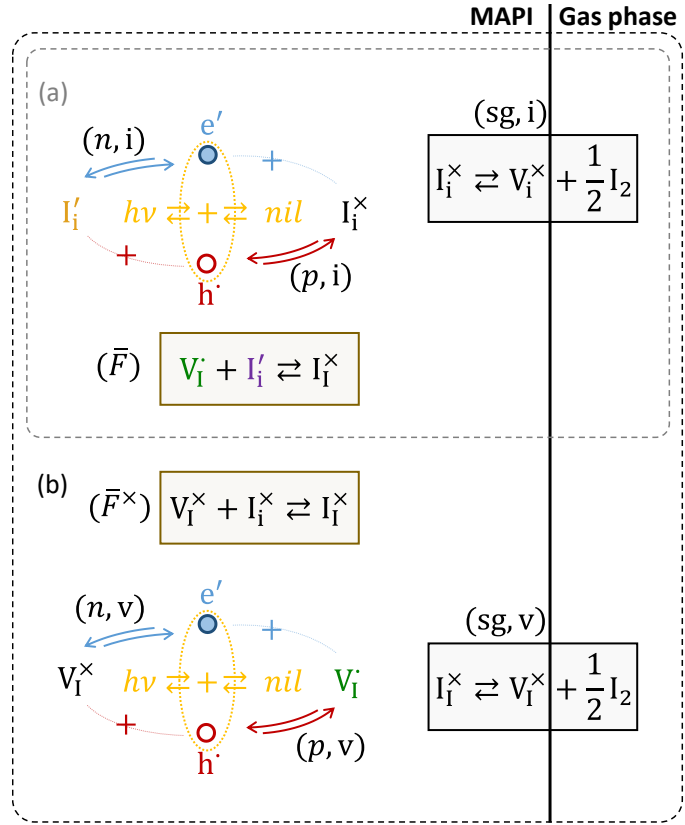

Figure 2. Schematics showing the coupling between optoelectronic reactions (generation and recombination) and the iodine (quasi)equilibrium of MAPI with the gas phase (sg is short for solid-gas). (a) Only iodide and iodine interstitials are redox-active ionic defects. (b) Both vacancies and interstitials are redox-active defects.

***Electronic properties.*** The electronic generation-recombination reaction involves multiple pathways[31] which can be described by means of net recombination terms $U_k = R_k - G_{th,k}$, where $R_k$ and $G_{th,k}$ are the recombination and thermal generation rates associated with process $k$. The following expressions describe the radiative ($k = rad$), Shockley-Read-Hall ($k = SRH$) and the Auger ($k = Aug$) processes:

$$U_{rad} = k_{rad}(np - n_i^2) \quad (4)$$

$$U_{SRH} = \frac{np - n_i^2}{\tau_n(p + p_1) + \tau_p(n + n_1)} \quad (5)$$

$$U_{Aug} = \gamma_n(n^2 p - n_{eq}^2 p_{eq}) + \gamma_p(np^2 - n_{eq} p_{eq}^2) \quad (6)$$

Here, $k_{rad}$ is the radiative constant, $\tau_n$, $\tau_p$, $n_1$ and $p_1$ are the capture time constant for electrons and for holes and the concentration parameters describing the trap energy position, $\gamma_n$ and $\gamma_p$ are the Auger coefficients. An additional term, $U_I$, is also considered, which refers to the net recombination rate deriving from the interaction of electrons and holes with the iodine defects (see Equation 12 below). In Table 1, the total generation rate $G$ is the sum of the term $G_{ext}$, which corresponds to the external light bias, and all thermal generation terms, while the total recombination rate $R$ is the sum of the rates associated with each recombination process.

*Ionic properties.* For the ionic situation, we concentrate on anti-Frenkel disorder[32–34] involving iodide vacancies and iodide interstitials (see also energy diagrams in Figure 1). Being aware of the fact that Schottky disorder (iodine vacancies and methylammonium vacancies $V_I^{\cdot}$ and $V_{MA}'$) is most likely the dominant ionic disorder in MAPI,[34,8] we assume anti-Frenkel disorder to be dominant for the purpose of referring to a simple and straightforward model. The general conclusions of this study can be applied to the Schottky disorder case too.

Although the ionic defect concentrations are not directly influenced by illumination, the coupling via defect chemical reactions leads to a situation of quasi-equilibrium for the ionic situation, too. In this case (as for any other bias) a quasi-electrochemical potential can be assigned to each ionic defect. Besides the mass-action constant associated with the anti-Frenkel disorder reaction at equilibrium, $K_{\bar{F}}$, we define a chemical potential of the anti-Frenkel ionic defects $\Delta\mu_{\bar{F}}$ to describe the nonequilibrium case, in analogy with $QFLS$ for the electronic charges:

$$\Delta\mu_{\bar{F}} = \delta\tilde{\mu}_{I_i'} + \delta\tilde{\mu}_{V_I^{\cdot}}. \quad (7)$$

In section (a) of the Results, we discuss under what conditions an increase in electronic concentrations induces a shift in the ionic anti-Frenkel equilibrium, involving changes in the values of $[I_i']$ and $[V_I^{\cdot}]$, while maintaining $[V_I^{\cdot}][I_i'] = K_{\bar{F}}$, $\Delta\mu_{\bar{F}} = 0$ and $-\tilde{\mu}_{V_I^{\cdot}}^* = \tilde{\mu}_{I_i'}^* = \tilde{\mu}_{I^-}^*$. Other situations where the ionic disorder is taken out-of-equilibrium ($\Delta\mu_{\bar{F}} \neq 0$) are explored in section (b).

Finally, based on reaction ($\bar{F}$) shown in Table 1, we introduce the net-recombination term $U_{\bar{F}}$ for anti-Frenkel defect pairs. Assuming $U_{\bar{F}}$ to follow a bimolecular process, we write

$$U_{\bar{F}} = \overleftarrow{k}_{\bar{F}}[V_I^{\cdot}][I_i'] - \overrightarrow{k}_{\bar{F}} = \overleftarrow{k}_{\bar{F}}([V_I^{\cdot}][I_i'] - K_{\bar{F}}), \quad (8)$$

where $\overleftarrow{k}_{\bar{F}}$ and $\overrightarrow{k}_{\bar{F}}$ are the rate constants for the recombination and thermal generation of anti-Frenkel defect pairs. The term $\overleftarrow{k}_{\bar{F}}$ can be related with the recombination Onsager radius $r_{\bar{F}}$ describing the recombination of (oppositely) charged defects as:

$$\overleftarrow{k}_{\bar{F}} = 4\pi D_{\bar{F}} r_{\bar{F}} \quad (9)$$

where $D_{\bar{F}}$ is the sum of the diffusion coefficients of the two defects participating in the recombination process.[35,36]

The analogous treatment of the neutral iodine defects is included in Table 1 together with the definition of parameters $K_{\bar{F}\times}$, $\Delta\mu_{\bar{F}\times}$. The net-recombination $U_{\bar{F}\times}$ can be defined similarly to Equation 8 (including $\overleftarrow{k}_{\bar{F}\times} = 4\pi D_{\bar{F}\times} r_{\bar{F}\times}$). In this work, we consider the solid-gas iodine exchange reaction as the only channel through which neutral iodine defects interact. This implies a net-recombination term of neutral defects

$$U_{\bar{F}\times} = \overrightarrow{R}_{sg,i} - \overleftarrow{R}_{sg,i} = \overleftarrow{R}_{sg,v} - \overrightarrow{R}_{sg,v} \quad (10)$$

A contribution describing reactions of neutral defects in the bulk of the mixed conductor (reaction ($\bar{F}^\times$) in Table 1, also shown in Figure 2b) would be necessary when looking at other situations (*e.g.* encapsulated samples).

*Iodine exchange and redox reactions.* At equilibrium, the following reaction can be used to describe the iodine exchange between MAPI and the gas phase.

$$I_I^\times + h^\cdot \rightleftharpoons \tfrac{1}{2} I_2, g + V_I^\cdot \quad (11)$$

Under light, electronic and ionic defects are no longer (necessarily) at equilibrium. Reaction (11) is therefore not sufficient to evaluate the quasi-equilibrium situation, and a kinetic model that includes all relevant redox reactions and iodine exchange reactions is required. We describe the interaction between ionic and electronic defects and the exchange of iodine at the solid-gas interface focusing on redox reactions where iodide interstitials or iodide vacancies interacting with either electrons or holes yield $I_i^\times$ or $V_I^\times$. Iodine in the gas phase is then either released from MAPI or incorporated in the structure via interaction with such neutral defects (Table 1 and Figure 2). In the main text of this study, the neutral defects are assumed to be at equilibrium with each other and with the gas phase.

A key aspect in the discussion of how the redox reactions influence the defect concentrations profiles concerns which of the reactions mediated by electrons or holes is dominant. To parameterize such aspect, we introduce the parameters $\Gamma_{I,i} = \vec{R}_{p,i}/\vec{R}_{n,i}$ and $\Gamma_{I,v} = \vec{R}_{p,v}/\vec{R}_{n,v}$ for reactions involving interstitials and vacancies, respectively, to express the extent to which each reaction is more favorable when mediated by conduction band electrons or by valence band holes. The $\Gamma_{I,i}$ and $\Gamma_{I,v}$ parameters are evaluated in the equilibrium case and at the intrinsic composition condition, *i.e.* $n_{eq} = p_{eq} = n_i$ and $[V_I^\cdot]_{eq} = [I_i']_{eq} = K_{\bar{F}}^{1/2}$, occurring for $P(I_2) = P(I_2)_i$. Note that, at equilibrium, the $\Gamma_{I,i}$ and $\Gamma_{I,v}$ ratios are the same also when considering the backward reaction (*i.e.* $\vec{R}_{p,i}/\vec{R}_{n,i} = \overleftarrow{R}_{p,i}/\overleftarrow{R}_{n,i}$).

The trapping and release of electronic charge carriers from iodide defects can be treated using the Shockley-Read-Hall framework, as discussed above for $U_{SRH}$ mediated by an immobile trap (see section 2 of the Supporting Information). This also leads to an additional contribution to net recombination of electronic charge carriers $U_I$ (*e.g.* the combination of $I_i' + h^\cdot \to I_i^\times$ and $I_i^\times + e' \to I_i'$ corresponds to the recombination process $h^\cdot + e' \to$ nil; the same would follow if we consider the reactions involving vacancy defects).

We can express the steady-state net recombination contribution due to all the redox reactions involving iodine defects as

$$U_I = \sum_{w=i,v}(\overleftarrow{R}_{n,w} - \vec{R}_{n,w}) = \sum_{w=i,v}(\vec{R}_{p,w} - \overleftarrow{R}_{p,w}). \quad (12)$$

It is useful to parameterize the rates associated with mobile ion-mediated redox reactions with respect to other intrinsic rates. We define the parameters $\Gamma_{p,i}$ and $\Gamma_{n,v}$ as the normalized rate of hole trapping by an iodide interstitial and the normalized rate of electron trapping by an iodide vacancy, respectively. These rates are evaluated at equilibrium and at $P(I_2) = P(I_2)_i$, and they are normalized by the radiative recombination rate at the same condition (see Table 1 and section 2 of the Supporting Information).

*Steady-state solution.* The equations resulting from the model described above are shown in Table 2. They reflect the fact that, at steady-state, no net mass-exchange occurs, the net recombination of electronic charges compensates for the external generation term, and the net recombination of ionic defects due to redox reactions is equal and opposite to their net recombination due to the anti-Frenkel reaction. By also considering the condition of electroneutrality, a system of six equations is obtained.

The analytical treatment becomes involved, requiring Brouwer approximations and additional simplification of the recombination terms to extract simple expressions for the $P(I_2)$ dependence of the charge carrier concentrations. We present numerical solutions to the problem obtained by solving the equations in Table 2 using the MATLAB function 'fsolve'.

Table 2. Equations used to determine the non-equilibrium defect chemistry of MAPI, based on the reactions shown in Figure 2 and Table 1.

| | |
|---|---|
| *1. Electronic quasi-equilibrium* | $G_{ext} = U_{rad} + U_{SRH} + U_{Aug} + U_I$ |
| *2. Electroneutrality* | $[V_I^\cdot] + [h^\cdot] - [I_i'] - [e'] = 0$ |
| *3. Iodide interstitials quasi-equilibrium* | $U_{\bar{F}} + \vec{R}_{n,i} - \overleftarrow{R}_{n,i} + \vec{R}_{p,i} - \overleftarrow{R}_{p,i} = 0$ |
| *4. Iodide vacancies quasi-equilibrium* | $U_{\bar{F}} - \vec{R}_{n,v} + \overleftarrow{R}_{n,v} - \vec{R}_{p,v} + \overleftarrow{R}_{p,v} = 0$ |
| *5. Iodine interstitials quasi-equilibrium* | $\vec{R}_{n,i} - \overleftarrow{R}_{n,i} + \vec{R}_{p,i} - \overleftarrow{R}_{p,i} - \vec{R}_{sg,i} + \overleftarrow{R}_{sg,i} = 0$ |
| *6. Iodine vacancies quasi-equilibrium* | $-\vec{R}_{n,v} + \overleftarrow{R}_{n,v} - \vec{R}_{p,v} + \overleftarrow{R}_{p,v} + \vec{R}_{sg,v} - \overleftarrow{R}_{sg,v} = 0$ |

## Results and discussion

### a. Single redox-active mobile ionic defect: ionic disorder at equilibrium

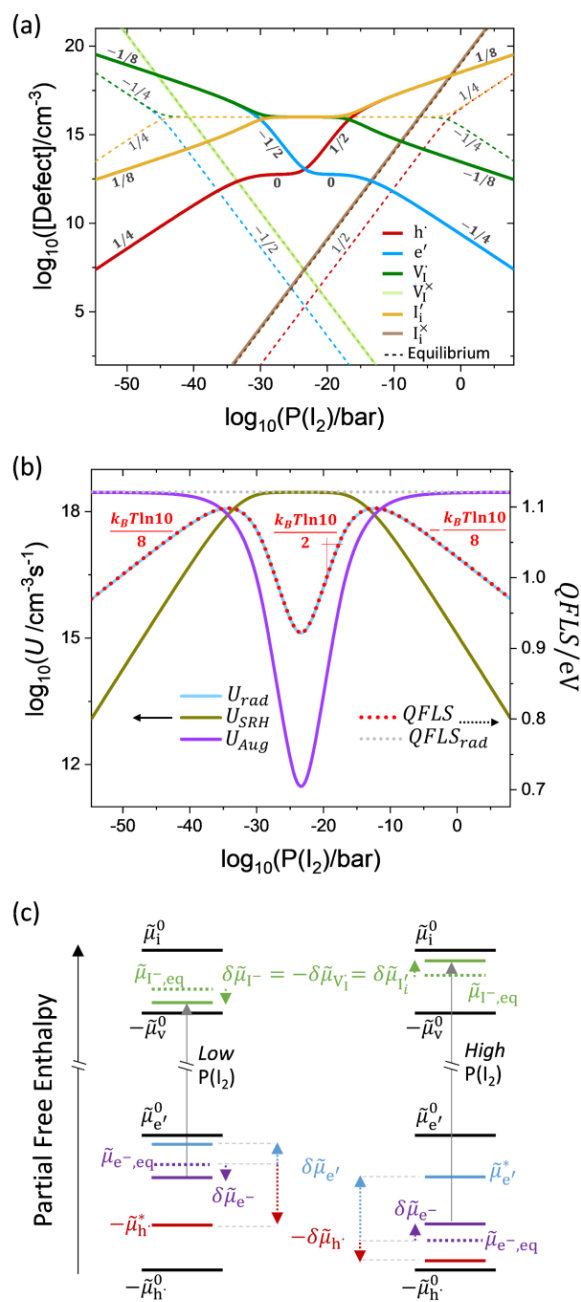

Figure 3. (a) Iodine partial pressure ($P(I_2)$) dependence of the steady-state electronic and ionic defect concentrations in MAPI at equilibrium and under light ($\sim 10^{-3}$ suns) plotted in a Kröger-Vink diagram (the numbers close to the data refer to the corresponding slopes). The calculation refers to assuming redox reaction occurring with iodine interstitials only, and $\Gamma_{I,i} = 1$. (b) Net recombination contributions and quasi-Fermi level splitting. (c) Schematic energy diagram emphasizing the change in position of the electrochemical potentials with respect to the electronic and ionic standard potentials, going from equilibrium (dashed lines) to the situation under light (solid lines), and for cases of low and high $P(I_2)$.[8]

We start by considering scenario (a) in Figure 2, where electronic charge carriers interact only with the interstitial defects $I_i^\times/I_i'$, while the rate constants associated with the $V_I^\times/V_I^\cdot$ reactions are negligibly small. Figure 3a shows the ionic and electronic defect concentrations as a function of $P(I_2)$ for the equilibrium (dashed lines) and the situation under light (solid lines). We consider the (symmetrical) case of $\Gamma_{I,i} = 1$ (comparable rates for the redox reactions mediated by electrons or holes at $P(I_2)_i$), and constant optical excitation of $\sim 10^{-3}$ suns equivalent. As expected, the electronic charge concentration is always larger under illumination than at equilibrium. Such increase is accompanied by a narrowing of the intrinsic region which is the $P(I_2)$ range where ionic defects are larger in concentration than electronic defects.

The concentration of iodide defects is perturbed under light too, as shown by the trend of $[V_I^\cdot]$ and $[I_i']$ deviating from the equilibrium profile, especially at low or high $P(I_2)$. Based on electroneutrality, a significant increase in $[V_I^\cdot]$ with respect to equilibrium at low $P(I_2)$ compensates for the large concentration of electrons obtained under light (see also $[I_i']$ and hole concentration at high $P(I_2)$). Interestingly, the profiles of $[V_I^\cdot]$ and $[I_i']$ still obey the anti-Frenkel equilibrium even under light, that is while their individual values are different from the equilibrium case, their product still corresponds to $K_{\bar{F}}$. The concentration of the neutral ionic defects $[I_i^\times]$ and $[V_I^\times]$ vary with $P(I_2)$ according to the mass-action laws for the solid-gas exchange reactions (sg, i and sg, v in Table 1), whether the system is at equilibrium or under light (their profiles are omitted in all figures below).

The data illustrate that, depending on $P(I_2)$, illumination induces a net iodine uptake from or release to the gas phase compared to the equilibrium condition. This results in a steady-state stoichiometry ($\delta^*$ in $\text{MAPbI}_{3+\delta^*}$, $\delta^* \propto [I_i'] + [I_i^\times] - [V_I^\cdot] - [V_I^\times]$) that differs from the equilibrium stoichiometry ($\delta_{eq}$ in $\text{MAPbI}_{3+\delta_{eq}}$). Importantly, we are ignoring any formation of higher order defects due to photo-generated electronic charges,[14] meaning that changes in ionic defect concentrations are expected under bias regardless of such events occurring. Note that such stoichiometry changes may correspond to compositions outside of the material's stability region, an aspect that is not included here.[14]

We conclude that, if only one ionic defect is involved in redox reactions (here interstitials $I_i^\times/I_i'$, but similar results would be obtained for vacancies), the application of a light bias to the mixed conductor effects the following:

- The concentration of the neutral iodine defects $[I_i^\times]$ and $[V_I^\times]$ are exclusively determined and fixed by the value of $P(I_2)$, *i.e.* light-independent, based on the mass-action constants $K_{sg,i}$ and $K_{sg,v}$ in Table 1.
- The rate equations involving reactions $(n, i)$ and $(p, i)$, and the generation and recombination of electronic charge carriers determine the values of $n$, $p$ and $[I_i']$.
- The value of $[V_I^\cdot]$ is fixed based solely on the value of $[I_i']$ according to the anti-Frenkel disorder reaction, which may shift but remains in equilibrium ($\Delta \mu_{\bar{F}} = 0$). Such shift corresponds to a change in stoichiometry with respect to the equilibrium condition.

Essentially, to obtain the data in Figure 3, only equations 1–4 in Table 2 need to be solved in combination with the mass-action laws for reactions ($\bar{F}$), (sg, i) and (sg, v). For all calculations in this section, we can obtain exact solutions to the problem by referring to the assumption that neutral defects remain in equilibrium with the gas phase, a condition that we refer to below as "*sg-eq*".

In Figure 3b, we display the resulting graphs of $QFLS$ as function of $P(I_2)$, as well as of the rates of recombination associated with the different mechanisms considered here. We find that, $U_{SRH}$ dominates for $P(I_2) \approx P(I_2)_i$, while $U_{Aug}$ is dominant in the N and P regions (very high or very low $P(I_2)$). The $QFLS$ is always lower than for the radiative limit ($QFLS_{rad}$) and it approaches such limit for two narrow ranges of $P(I_2)$. The position of the $QFLS$ maxima correspond to the situations where the contribution of $R_{rad}$ to the total recombination is fractionally largest, and where $R_{SRH} = R_{Aug}$. Note that the latter condition is not general, but it is based on the input parameters used here ($\tau_n = \tau_p$, $n_1 = p_1$ and $\gamma_n = \gamma_p$). The local minimum observed for $n = p$ is consistent with the Shockley-Read-Hall theory of recombination (for the case of $\tau_n = \tau_p$ and for a mid-gap trap level). The slope of the $QFLS$ vs $P(I_2)$ can be evaluated as the sum of the slopes associated with electrons and holes in the Kröger-Vink diagram times a $k_B T \ln[10]$ factor. The data in Figure 3b highlight that the $QFLS$ obtained for a mixed conducting film (here MAPI) depends non-monotonically on the component partial pressure (here $P(I_2)$). Because $QFLS$ is a proxy for the maximum open circuit potential which can be achieved once the material is embedded in a complete solar cell, this analysis emphasizes the importance of the component partial pressure for controlling the performance of optoelectronic devices based on mixed conductors.[37]

In Figure 3c, we illustrate schematically the corresponding changes in electrochemical potentials for all defects where changes in both the ionic and electronic situation under bias (solid lines) are compared with the equilibrium solution (dashed lines). Consistently with the data in Figure 3a, $\delta\tilde{\mu}_{V_I} = -\delta\tilde{\mu}_{I_i'} = -\delta\tilde{\mu}_{I^-}$ ($\Delta\mu_{\bar{F}} = 0$) at any $P(I_2)$, with $\delta\tilde{\mu}_{V_I} \approx 0$ in the intrinsic region under light. Because the steady-state condition $\mu_{I,MAPI} = \frac{1}{2}\mu_{I_2,g}$ is valid both at equilibrium and under bias, based on $\mu_{I,MAPI} = \tilde{\mu}_{I^-}^*, -\tilde{\mu}_{e^-}^*$, it follows that $\delta\tilde{\mu}_{I^-} = \delta\tilde{\mu}_{e^-}$ at any given $P(I_2)$. In section 3 of the Supporting Information, we provide more details for the analysis of this quasi-equilibrium as well as the asymptotic trends for very low and very high $P(I_2)$.

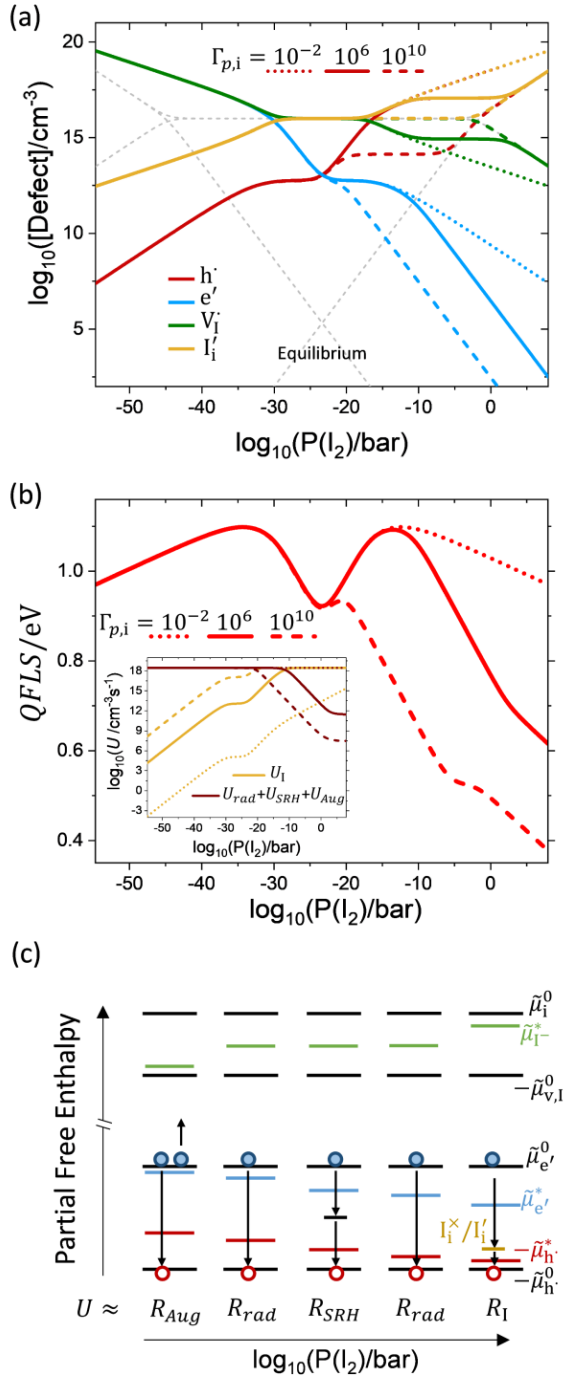

Figure 4. (a) Defect concentration and (b) $QFLS$ for a MAPI film calculated for $\Gamma_{p,i} = 10^{-2}, 10^6$ and $10^{10}$ (corresponding to $\vec{k}_{p,i} = 2.1 \times 10^{-24}, 2.1 \times 10^{-16}$ and $2.1 \times 10^{-12}\ cm^3 s^{-1}$). The inset in (b) shows the comparison between the combined recombination contribution from radiative, SRH and Auger mechanisms, and the contribution $U_I$ due to the redox reactions (n,i) and (p,i). Illumination of $10^{-3}$ suns and $\Gamma_{I,i} = 1$ are considered. (c) Generalized energy diagram showing the dominant recombination mechanisms for different $P(I_2)$ regions in (a) and (b) ($\Gamma_{p,i} = 10^6$ case). The $I_i^\times / I_i'$ energy level is included (~0.3 eV above the valence band maximum based on input parameters).

The data in Figure 3 refer to the case of $\Gamma_{p,i} = 10^{-2}$ (low hole trapping rate by $I_i'$). Combined with a balanced interaction of interstitials with electrons and holes ($\Gamma_{I,i} = 1$), this corresponds to a negligible contribution of recombination due to redox reactions with iodide interstitials $U_I$ compared to other recombination pathways for the selected light intensity. In Figure 4, we display the calculated defect concentrations and $QFLS$ using the same input parameters as in Figure 3, but with varying rate of hole trapping at iodine interstitials, parameterized through $\Gamma_{p,i}$. Simultaneous relative variations in electron trapping at iodide interstitials are ensured by selecting $\Gamma_{I,i} = 1$ in all cases. Figure 4a indicates that, for large values of $\Gamma_{p,i}$, all defect concentrations are varied significantly from the situation shown in Figure 3 in the high $P(I_2)$ region. The resulting $QFLS$ (Figure 4b) shows a significant drop in such region, which is ascribed to a dominant electron-hole recombination contribution mediated by the forward reactions associated with $(n, i)$ and $(p, i)$. Such contribution is compared with the total net recombination deriving from all the other mechanisms considered (inset of Figure 4b).

We note that the Shockley-Read-Hall rate for recombination mediated by immobile traps is established based on the parameters $n_1, p_1$, which depend on the energetic position of the trap, and $\tau_n, \tau_p$, which depend on the (fixed) concentration and the capture coefficient of the trap. While the recombination due to redox reactions involving $I_i^\times / I_i'$ follows a similar principle, the concentration of such recombination centers is determined by the overall evaluation of the charge carrier quasi-equilibrium. In Figure 4, $U_I$ is dominant at high $P(I_2)$, due to the increase in $I_i^\times$ and $I_i'$ defect concentrations. The dominant recombination mechanism for different $P(I_2)$ values is shown schematically in Figure 4c (referring to the $\Gamma_{p,i} = 10^6$ case). The $I_i^\times / I_i'$ energy level is included within the energy bandgap.

Figure 5a shows the influence of $\Gamma_{I,i}$ on the defect quasi-equilibrium. As reaction (sg,i) is at equilibrium in this scenario, $\Gamma_{I,i}$ is a measure of the degree to which the "hole channel" vs the "electron channel" control the iodine incorporation/excorporation at equilibrium (see Table 1 and Figure 2). Changing $\Gamma_{I,i}$ does not vary the equilibrium defect concentrations (dashed lines in the Kröger-Vink diagrams) as these depend only on the mass-action constants. On the other hand, the position of the intrinsic region under light shifts on the $P(I_2)$ axis when varying $\Gamma_{I,i}$, while always remaining within the boundaries of the intrinsic region defined by the equilibrium case ($n \geq n_{eq}$ and $p \geq p_{eq}$). The iodine partial pressure $P(I_2)_i^*$ at which $n = p$ and $[V_I^{\cdot}] = [I_i']$ refers to the intrinsic condition under light. While $P(I_2)_i^*$ is essentially the same as $P(I_2)_i$ for the example in Figure 3 ($\Gamma_{I,i} = 1$ and low $\Gamma_{p,i}$), Figure 5a shows that, in general, $P(I_2)_i^* \neq P(I_2)_i$.

Figure 5b highlights the shrinking of the intrinsic region with increasing light intensity (the same input parameters as for Figure 3 are used). The trends in $QFLS$ resulting from varying $\Gamma_{I,i}$ or light intensity are shown in Figure 5c, emphasizing that the $P(I_2)$ values corresponding to local $QFLS$ maxima are dependent on such parameters. The situation is further complicated if larger values of $\Gamma_{p,i}$ are considered (in Figure 5, $\Gamma_{p,i} = 10^{-2}$). Figure 5d displays the light intensity dependence of $QFLS$ evaluated at different values of $P(I_2)$ for $\Gamma_{I,i} = 1$. The data once again illustrate the influence of $P(I_2)$ on the dominant recombination mechanism, as also highlighted by the trends in local ideality factor (defined here as $\frac{1}{k_B T} \frac{d(QFLS)}{d\ln(Intensity)}$, see inset).

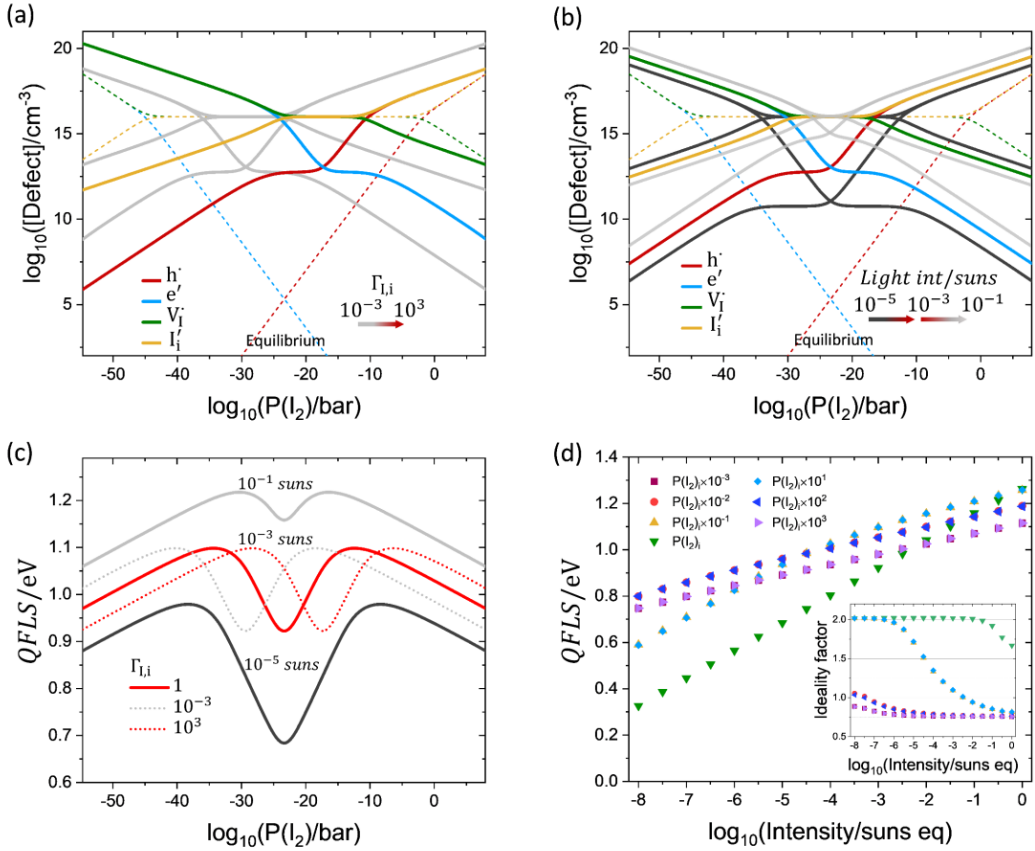

Figure 5. Calculated defect concentrations vs $P(I_2)$ in MAPI (a) under $\sim 10^{-3}$ suns equivalent illumination comparing situations where $\Gamma_{I,i} = 10^{-3}$ or $\Gamma_{I,i} = 10^3$ and (b) for $\Gamma_{I,1} = 1$ for varying bias light intensities ($\Gamma_{p,i} = 10^{-2}$ in all cases). The corresponding $QFLS$ profiles for the data in (a) and (b) are shown in (c). (d) Ideality factor analysis obtained from the light intensity dependent calculations of the $QFLS$.

### b. Multiple redox-active mobile ionic defects: ionic disorder out of equilibrium

We now consider situation (b) in the model shown in Figure 2, whereby both interstitial and vacancy defects interact with electronic charge carriers via redox reactions. For now, we shall continue to assume the "*sg-eq*" condition. Such a situation may be relevant only to very small particles of the mixed conductor (although see discussion below).

Figure S2 shows that if both interstitials and vacancies show similar coupling to electrons and to holes ($\Gamma_{I,i} = \Gamma_{I,v}$) and the recombination mediated by the mobile ions is negligible (low $\Gamma_{p,i}$ and $\Gamma_{n,v}$), essentially unchanged trends are found compared with the results obtained using a single redox-active ion (Figure 3 and 5). If either of these conditions are not met, different trends from the single redox active mobile defect case are obtained (see Figure S3 for $\Gamma_{I,i} \neq \Gamma_{I,v}$ and low $\Gamma_{p,i}$, $\Gamma_{n,v}$, and Figure S4 for the case of high $\Gamma_{p,i}$, $\Gamma_{n,v}$ and $\Gamma_{I,i} = \Gamma_{I,v}$). Despite such deviations, equilibrium in the ionic disorder is maintained under light, although shifted with respect to the dark equilibrium case.

We now discuss the implications of both above conditions not being met. Figure 6 explores the effect of the parameters $\Gamma_{I,i}$ and $\Gamma_{I,v}$, which define to what extent the $I_i^\times/I_i'$ and the $V_I^\times/V_I^\cdot$ quasi-equilibria are established by reactions with holes ($\Gamma_I \gg 1$) or electrons ($\Gamma_I \ll 1$). The data are obtained considering a value of $\Gamma_{p,i} = \Gamma_{n,v}$ as large as $10^{11}$.

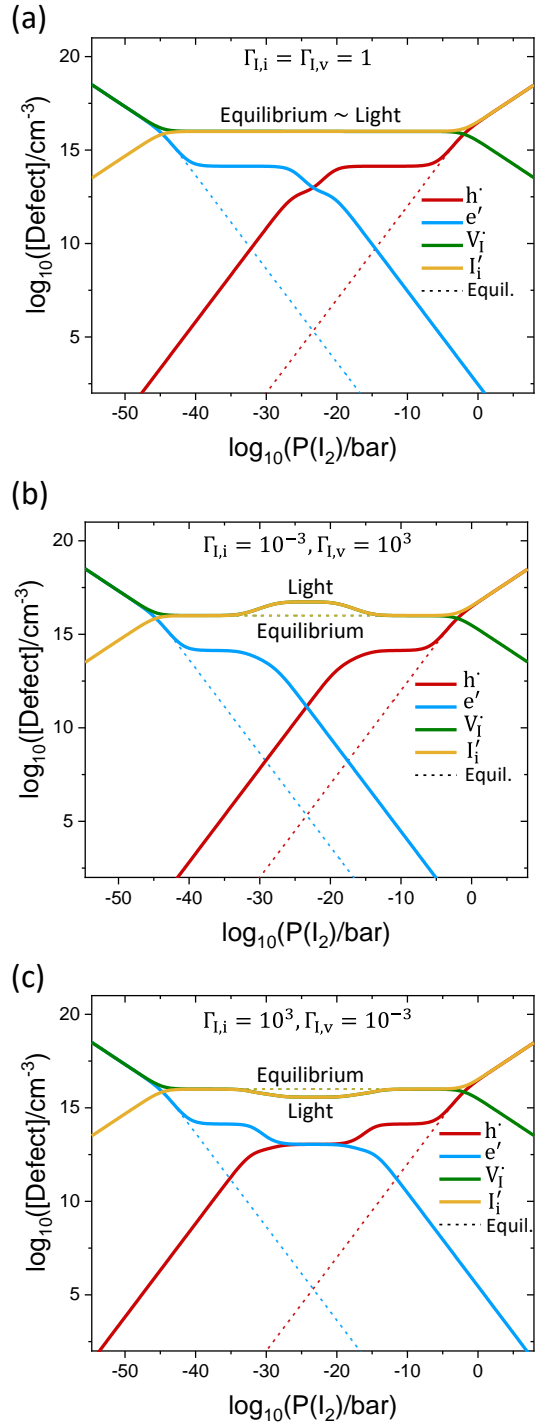

Figure 6. Kröger-Vink diagrams for a mixed conductor (MAPI) at equilibrium and under light (10$^{-3}$ sun equivalent illumination), for situations where both iodide vacancies and interstitials participate in redox reactions with electrons or holes. Different values of the parameters $\Gamma_{I,i}$ and $\Gamma_{I,v}$ are used for the case of $\Gamma_{p,i} = \Gamma_{n,v} = 10^{11}$. (a) $\Gamma_{I,i} = \Gamma_{I,v} = 1$; (b) $\Gamma_{I,i} = 10^{-3}, \Gamma_{I,v} = 10^3$; (c) $\Gamma_{I,i} = 10^3, \Gamma_{I,v} = 10^{-3}$.

Figure 6a is the reference situation where the redox reactions involving interstitials and the ones involving vacancies are driven by electrons and holes to a similar extent ($\Gamma_{I,i} = \Gamma_{I,v}$), as discussed above for Figure S4. As $\Gamma_{I,i} = \Gamma_{I,v} = 1$ (and $\Gamma_{p,i} = \Gamma_{n,v}$), the defect diagram preserves "symmetry" about the pressure value $P(I_2)_i$. The ionic defect profiles under light largely follow the same trends in the equilibrium and in the light biased cases. This contrasts with the results in Figure 3, where significant changes in ionic defect concentrations are encountered especially in the high and low $P(I_2)$ ranges. Due to the large values of the redox reactions' rate constants (related to $\Gamma_{p,i}$ and $\Gamma_{n,v}$), the iodide vacancy reduction reaction $(n, v)$ at low $P(I_2)$ and the iodide interstitial oxidation $(p, i)$ at high $P(I_2)$ are essentially operating at equilibrium. Along with the *sg-eq* condition, this ensures minimal deviations from the equilibrium trends of the relevant defects in these two pressure ranges. The same occurs also for intermediate values of $P(I_2)$, due to the symmetrical interaction of the electronic carriers with the ionic defects.

Figures 6b and 6c consider situations where there is an imbalance in the relevance of electrons and holes in the determination of the redox reactions quasi-equilibrium, as expressed by the parameters $\Gamma_{I,i}$ and $\Gamma_{I,v}$, and such imbalance is not the same for interstitials and vacancies ($\Gamma_{I,i} \neq \Gamma_{I,v}$). The data show a striking increase or decrease (Figure 6b and 6c, respectively) in the concentration of both iodide vacancies and iodide interstitials under light with respect to equilibrium. Such observation implies a deviation from the anti-Frenkel equilibrium due to illumination, and it can be explained as follows. Increase in $n$ tends to increase $[I_i']$ but decrease $[V_I^\cdot]$, while an increase in $p$ has the opposite effect (see Table 1). Since $[I_i^\times]$ and $[V_I^\times]$ are fixed at any given $P(I_2)$, changes in the value of $[I_i']$ and $[V_I^\cdot]$ depend on the rates of the reactions that "connect" each of them to either electrons or holes. Such connection is parameterized through $\Gamma_{I,i}$ and $\Gamma_{I,v}$.

Analytically, it is useful to define the parameter

$$X_{vi} = (\overleftarrow{R}_{n,i} - \overrightarrow{R}_{n,i}) - (\overrightarrow{R}_{p,i} - \overleftarrow{R}_{p,i}) = (\overrightarrow{R}_{p,v} - \overleftarrow{R}_{p,v}) - (\overleftarrow{R}_{n,v} - \overrightarrow{R}_{n,v}). \qquad (13)$$

The absolute value of $X_{vi}$ corresponds to the absolute value of the electron-hole net recombination contribution involving two ionic defects. This can correspond to a positive net recombination where two ionic defects mediate the separate trapping of electrons and holes. It can also refer to a positive net thermal generation where the two defects mediate the separate de-trapping of electrons and holes.

If only one redox active defect is considered, $X_{vi} = 0$ by definition, at steady-state. If instead both interstitials and vacancies are redox active, as in Figure 6, such condition is no longer necessarily true. Specifically, if $X_{vi} = 0$, the rate of electron trapping by each of the iodide vacancies or iodine interstitials is counter-balanced by an equal rate of hole trapping interacting with the same type of defect (this is the case when $\Gamma_{I,i} = \Gamma_{I,v}$). That means that no reaction between ionic defects is required to mediate the recombination (or thermal generation process).

If instead $X_{vi} \neq 0$, the fraction of the total electron-hole net recombination which is mediated by separate (de-)trapping of electronic charges by the two ionic defects also requires reaction between such ionic defects. Specifically, at steady-state, such an electron-hole recombination process involves the recombination of the two ionic defects with the trapped carriers and the generation of the two original defects before trapping (similar argument applies to electron-hole thermal generation). With this in mind, we can rewrite the rate equations in Table 2 related to the ionic defects, as shown in Table

3. The effective ionic defect generation term for charged anti-Frenkel pairs ($G_{\bar{F}}$) and neutral anti-Frenkel pairs ($G_{\bar{F}\times}$) are defined and, at steady-state, they relate to each other and to the $X_{\text{vi}}$ term as follows:

$$G_{\bar{F}} = -G_{\bar{F}\times} = X_{\text{vi}} \qquad (14)$$

Importantly, and in contrast with the electronic generation term $G_{ext}$, these effective ionic defect generation terms can be positive or negative. This leads to the prediction that illumination can increase the concentration of ionic defects via a $G_{\bar{F}} > 0$ (analogously to the effect of a positive $G_{ext}$ on the electronic charge carrier concentrations) but also decrease the concentration of ionic defects via a $G_{\bar{F}} < 0$. Based on this interpretation, the trends in Figure 6 can be explained. By considering the parameter $h_{ion} = \log_{10}(\Gamma_{I,v}/\Gamma_{I,i})$, we find that situations where $G_{\bar{F}} = 0$, $G_{\bar{F}} > 0$ or $G_{\bar{F}} < 0$ refer to $h_{ion} = 0$, $h_{ion} > 0$ and $h_{ion} < 0$, respectively (see Figures 6a, b and c).

Table 3. Equations in Table 2, expressed based on the definition of the effective ionic defect generation terms $G_{\bar{F}}$ and $G_{\bar{F}\times}$ (see Equations 13 and 14).

| | |
|---|---|
| 1. Electronic quasi-equilibrium | $G_{ext} = U_{rad} + U_{SRH} + U_{Aug} + U_I$ |
| 2. Electroneutrality | $[V_I^·] + [h^·] - [I_i'] - [e'] = 0$ |
| 3. Iodide interstitial quasi-equilibrium | $G_{\bar{F}} = U_{\bar{F}}$ |
| 4. Iodide vacancies quasi-equilibrium | $G_{\bar{F}} = U_{\bar{F}}$ |
| 5. Iodine interstitial quasi-equilibrium | $G_{\bar{F}\times} = U_{\bar{F}\times}$ |
| 6. Iodine vacancies quasi-equilibrium | $G_{\bar{F}\times} = U_{\bar{F}\times}$ |

We note that a similar change in concentration compared with the equilibrium case affects both ionic defects in the intrinsic regions of Figure 6b and c. The resulting change in stoichiometry between the dark and the light-bias cases is therefore less significant here than for the situation displayed in the N and P regions of Figure 3a. While the latter largely reflects the discussion of previous studies on light-effects in STO,[19,22–24] obtaining simultaneous increase or decrease in the concentration of both ionic defects involved in the dominant ionic disorder reaction through light may open new opportunities in material science and beyond.

It is also important to note that the data in Figure 6 are obtained by considering the *sg-eq* condition. This means that the concentration profiles for the neutral defects are unperturbed under light compared with the equilibrium situation (see Figure 3a), and that only the concentration of the anti-Frenkel pairs is subject to variations, which depend on the value of $\Gamma_{I,i}$, $\Gamma_{I,v}$ and $\Gamma_{p,i}$, $\Gamma_{n,v}$. When solving the full kinetic model which includes a finite rate for the solid-gas exchange reactions (*sg-eq* condition no longer valid), we find that both the charged and the neutral ionic defect concentrations are perturbed by light. The effect described above, with either an increase or a decrease of both $[I_i']$ and $[V_I^·]$, still occurs but to a lesser extent (see section 5 of the Supporting Information).

Here, we discuss solutions under the *sg-eq* condition further, as these describe the upper limit to the anti-Frenkel nonequilibrium induced by illumination of the mixed-conductor. In Figure 7a, the $\Delta\mu_{\bar{F}}$ and the $QFLS$ profiles corresponding to the data in Figure 6 are illustrated. The conditions that lead to an enhancement in the ionic concentrations also lead to a drop in $QFLS$, while an increase in $n$ and $p$ is observed when redox reactions reduce the mobile ion concentrations. This can be explained since $U_I$ scales with the concentration of mobile ions that can mediate recombination. In Figure 7b, the electronic generation rate as well as the anti-Frenkel pair effective generation rate are shown.

Consistent with the definition of the effective ionic defect generation rate, $|G_{\bar{F}}| \leq G_{ext}$. Figure 7c shows a schematic including the extended energy level diagram summarizing these findings. In the Supporting Information we show other examples where the condition $\Gamma_{I,i} \neq \Gamma_{I,v}$ result in $\Delta\mu_{\bar{F}} \neq 0$, including the case where the same electronic charge carrier dominates redox reactions with both interstitials and vacancies (see Figure S6).

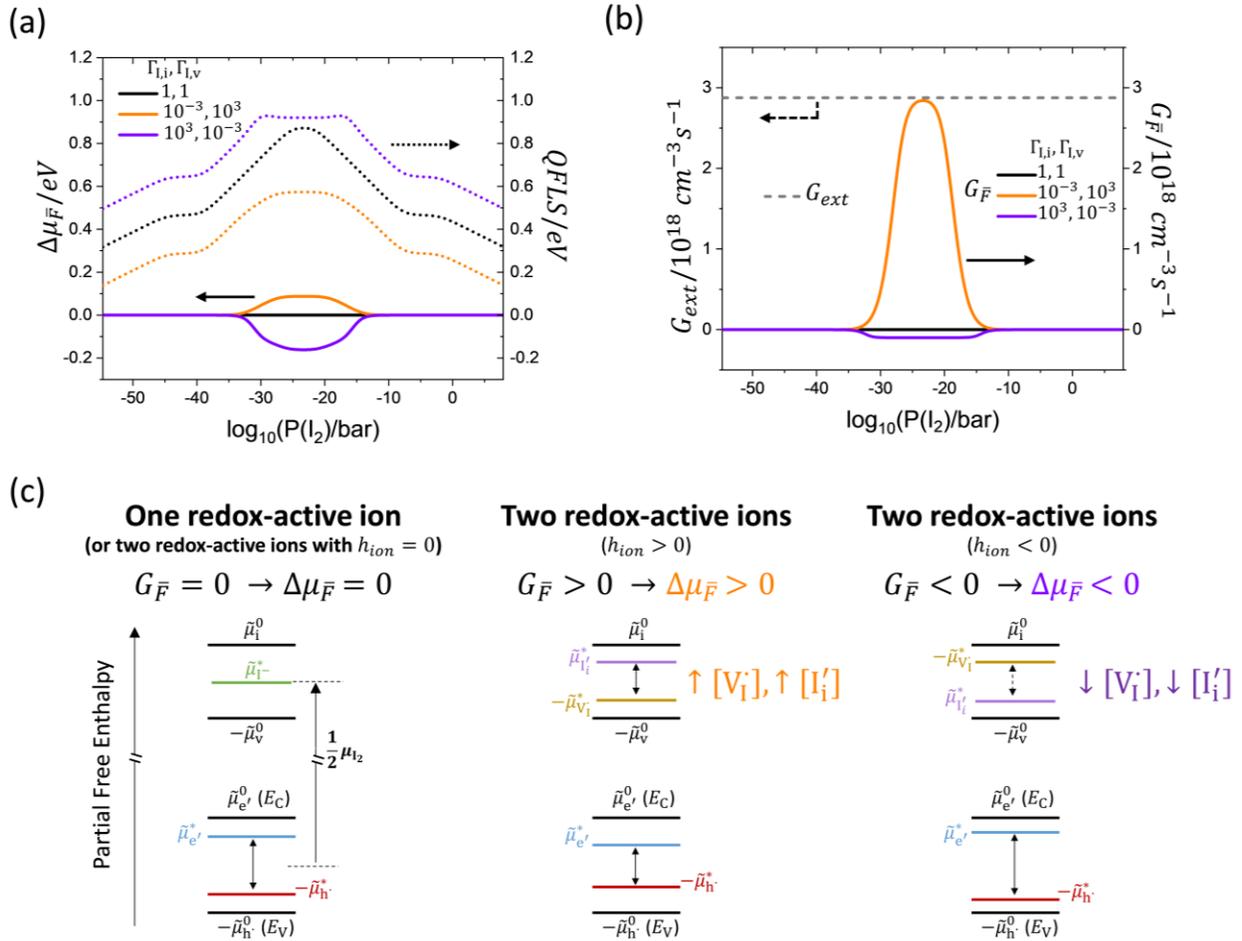

Figure 7. Consequences of including two redox-active mobile ionic defects in the defect chemical quasi-equilibrium of a mixed conductor exposed to light, assuming equilibrium of the neutral defects ($I_i^\times$ and $V_I^\times$) with the gas phase (*sg-eq*). (a) Ionic ($\Delta\mu_{\bar{F}}$) and electronic ($QFLS$) chemical potentials. (b) Electronic generation rate $G_{ext}$ used in the calculations ($10^{-3}$ suns equivalents) and resulting effective anti-Frenkel ionic generation rate $G_{\bar{F}}$ for the three situations considered in (a) and in Figure 6. (c) Generalized energy level diagram corresponding to the (left) unchanged, (center) increased and (right) decreased charged ionic defect concentration upon illumination.

Figure 8 illustrates the dependence of the ionic defect concentration enhancement or depression on the parameters $\Gamma_{I,i}$, $\Gamma_{I,v}$, and $\Gamma_{p,i}$, $\Gamma_{n,v}$, based on the trends in $\Delta\mu_{\bar{F}}$. While the sign as well as the magnitude of $\Delta\mu_{\bar{F}}$ depend on $h_{ion}$, the data show that obtaining perceptible deviations from the anti-Frenkel equilibrium requires a sufficiently large value of both $\Gamma_{p,i}$ and $\Gamma_{n,v}$. This is because the extent of the non-equilibrium in the anti-Frenkel disorder (magnitude of $\Delta\mu_{\bar{F}}$) correlates with the fraction of the overall electron-hole recombination rate that is due to mobile-ion mediated reactions (dictated by $\Gamma_{p,i}$, $\Gamma_{n,v}$ as well as by $\Gamma_{I,i}$, $\Gamma_{I,v}$).

Figure 8b and d highlight that, for increasing $\Gamma_{p,i}$ and $\Gamma_{n,v}$, the $QFLS$ decreases less substantially if $h_{ion} < 0$ than if $h_{ion} > 0$. This is due to the depression, rather than enhancement, in $[I_i']$ and $[V_I^·]$, which mediate recombination. In other words, whether recombination active ionic defects are being "pumped out of" or "pumped in" the mixed conductor by light significantly influences not only the ionic but also the electronic quasi-equilibrium. Section 7 of the Supporting Information shows similar calculations performed without the SRH recombination mediated by immobile defects to highlight the effect of minimizing the recombination due to anti-Frenkel pairs (for $h_{ion} < 0$) on the $QFLS$. We stress that such situation cannot improve the value of $QFLS$ beyond the limit where the anti-Frenkel defects are not recombination active (see Figure S7 and S8).

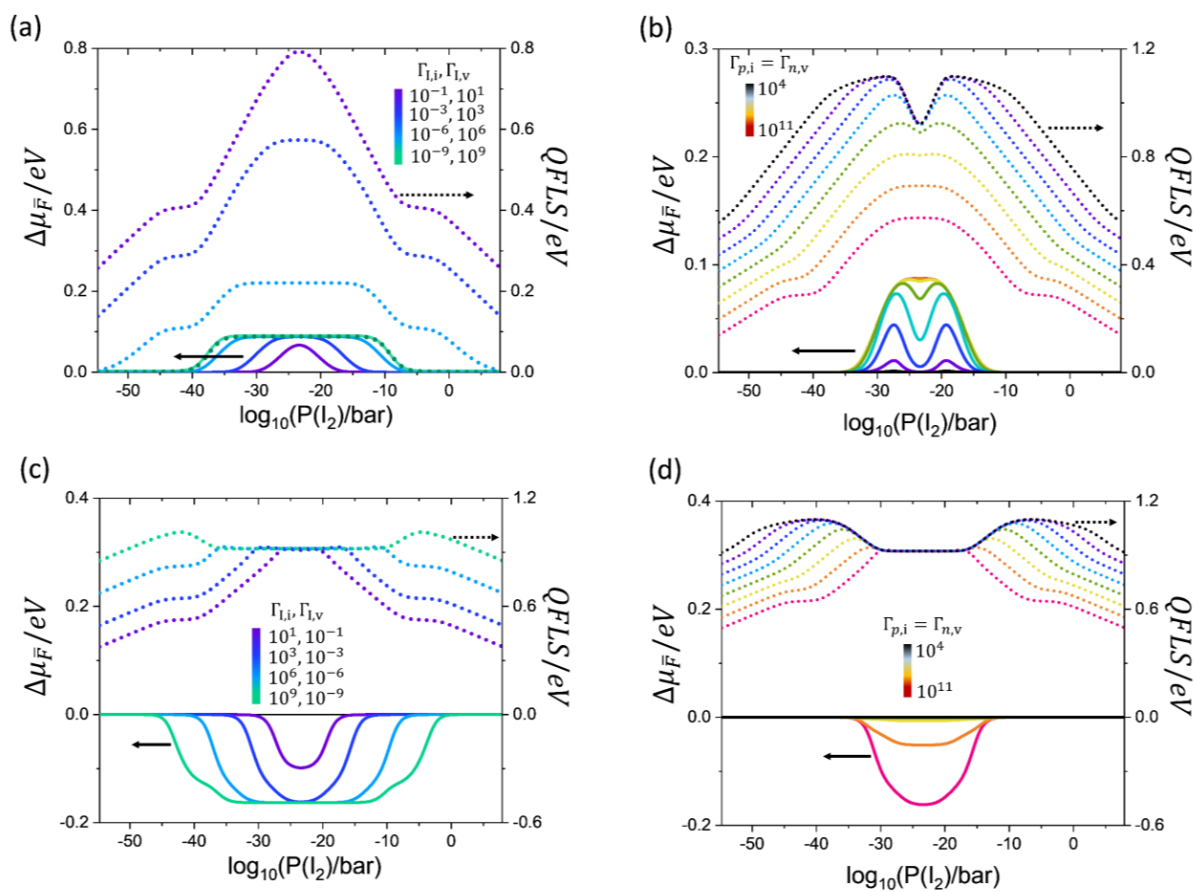

Figure 8. Dependence of the ionic ($\Delta\mu_{\bar{F}}$) and electronic ($QFLS$) change in chemical potential on the $\Gamma_{I,i}$, $\Gamma_{I,v}$, and $\Gamma_{p,i}$, $\Gamma_{n,v}$ parameters, as a function of P(I$_2$). Trends for (a) increasing $h_{ion} > 0$ ($\Gamma_{p,i} = \Gamma_{n,v} = 10^{11}$), (b) increasing $\Gamma_{p,i} = \Gamma_{n,v}$ ($\Gamma_{I,i} = 10^{-3}$, $\Gamma_{I,v} = 10^{3}$), (c) decreasing $h_{ion} < 0$ ($\Gamma_{p,i} = \Gamma_{n,v} = 10^{11}$), (d) increasing $\Gamma_{p,i} = \Gamma_{n,v}$ ($\Gamma_{I,i} = 10^{3}$, $\Gamma_{I,v} = 10^{-3}$). All remaining parameters are the same as the ones used in Figure 3. $h_{ion} = \log_{10}\left(\frac{\Gamma_{I,v}}{\Gamma_{I,i}}\right)$.

In summary, the analysis presented in this study:

- provides a framework for the study of redox reactions, electron-hole recombination and solid-gas component exchange reactions when photo-active mixed conducting materials are illuminated
- outlines guidelines for material design where ionic defect concentrations (trap densities) can be increased and, strikingly, also decreased by light
- proposes a rationale for the development of photoelectrochemical systems that allow for stoichiometry control using optical stimuli
- emphasizes the importance of understanding electronic and ionic kinetic properties, especially in the context of solar energy conversion devices based on mixed conductors where mobile ions provide a significant contribution to the total electron-hole recombination rate.

This work presents the fundamental effect that light has on the steady-state ionic and electronic bulk properties of mixed conductors. The model can be extended to include transport and interfacial effects, and it can be integrated in physical models of complete mixed conducting devices for the study of their steady-state as well as transient response when exposed to light and/or voltage bias.[20,29,38–42]

## Conclusions

We present a model that describes the electronic and ionic quasi-equilibria in a mixed ionic-electronic conductor exposed to light, using the halide perovskite methylammonium lead iodide (MAPI) as model system. We focus on the rate equations describing defect reactions and neglect defect transport to the solid-gas interface. By coupling the ionic (anti-Frenkel) disorder reaction, the iodine redox reactions and exchange with the gas phase with the electronic (photo-)generation and recombination, we predict the trends of ionic and electronic defect concentrations as function of the iodine chemical potential (partial pressure). We discuss the material's thermodynamic and kinetic parameters that control the steady-state solution to such problem, with particular focus on the relative contribution of the two electronic charge carriers to the iodine redox and exchange reactions. Knowledge of such aspects allows one to determine the steady-state quasi-Fermi level splitting in the material, with relevance to the optimization of solar cells. The equilibrium in the ionic disorder reaction is maintained and only shifted by light for situations where a single ionic defect is redox-active, regardless of the solid-gas exchange kinetics. Including redox reactions for both ionic defects (in this case iodide vacancies and interstitials) opens the possibility of taking the ionic disorder reaction out-of-equilibrium. This points to the intriguing opportunity of designing systems where defect concentrations are increased or decreased by light ("light-driven ionic defect pump"). Such scenarios are described by introducing an ionic defect pair chemical potential and an effective ionic defect generation rate, both of which can be positive (in analogy to their electronic counter-parts) but also negative, depending on the ionic-electronic interactions.

## Acknowledgements

The authors are grateful to Rotraut Merkle for reading and providing useful feedback on this manuscript. D.M. acknowledges financial support by the Alexander von Humboldt Foundation.

## References

1   J. Maier, *Physical Chemistry of Ionic Materials*, Wiley-VCH Verlag GmbH & Co. KGaA, 2nd edn., 2023.
2   S. M. Sze and K. N. Kwok, *Physics of Semiconductor Devices Physics of Semiconductor Devices*, John Wiley & Sons, Inc., Hoboken, New Jersey, 1995, vol. 10.

3   P. Y. Yu and M. Cadorna, *Fundamentals of Semiconductors*, Springer, 4th edn., 2008.
4   J. Nelson, *The Physics of Solar Cells*, Imperial College Press, 2003.
5   P. Würfel, *Wiley-VCH Verlag GmbH & Co. KGaA*, 2005, preprint.
6   F. A. Kroeger, *Chemistry of Imperfect Crystals*, North-Holland, Amsterdam, 1964.
7   J. Maier, *Angewandte Chemie International Edition In English*, 1993, **32**, 313–335.
8   D. Moia and J. Maier, *Materials Horizons*, 2023, **10**, 1641–1650.
9   T. Y. Yang, G. Gregori, N. Pellet, M. Grätzel and J. Maier, *Angewandte Chemie - International Edition*, 2015, **54**, 7905–7910.
10  C. Eames, J. M. Frost, P. R. F. Barnes, B. C. O'Regan, A. Walsh and M. S. Islam, *Nature Communications*, 2015, **6**, 2–9.
11  A. Senocrate, I. Moudrakovski, G. Y. Kim, T.-Y. Yang, G. Gregori, M. Grätzel and J. Maier, *Angewandte Chemie International Edition*, 2017, **56**, 7755–7759.
12  N. Leupold, A. L. Seibel, R. Moos and F. Panzer, *European Journal of Inorganic Chemistry*, 2021, **2021**, 2882–2889.
13  E. Ghahremanirad, S. Olyaee and J. Bisquert, *Journal of Physical Chemistry Letters*, 2017, 1402–1406.
14  G. Y. Kim, A. Senocrate, T. Yang, G. Gregori, M. Grätzel and J. Maier, *Nature Materials*, 2018, **17**, 445–450.
15  D. Moia, I. Gelmetti, M. Stringer, O. Game, D. Lidzey, E. Palomares, P. Calado, J. Nelson, W. Fisher and P. R. F. Barnes, *Energy & Environmental Science*, 2019, **12**, 1296–1308.
16  D. A. Egger, A. Bera, D. Cahen, G. Hodes, T. Kirchartz, L. Kronik, R. Lovrincic, A. M. Rappe, D. R. Reichman and O. Yaffe, *Advanced Materials*, 2018, **30**, 1–11.
17  Y. Zhou, L. You, S. Wang, Z. Ku, H. Fan, D. Schmidt, A. Rusydi, L. Chang, L. Wang, P. Ren, L. Chen, G. Yuan, L. Chen and J. Wang, *Nature Communications*, 2016, **7**, 1–8.
18  Y. Rakita, I. Lubomirsky and D. Cahen, *Mater. Horiz.*, 2019, **6**, 1297–1305.
19  R. Merkle and J. Maier, *Physical Chemistry Chemical Physics*, 2002, **4**, 4140–4148.
20  A. Senocrate, E. Kotomin and J. Maier, *Helvetica Chimica Acta*, 2020, **103**, 1–8.
21  T. Defferriere, D. Klotz, J. C. Gonzalez-Rosillo, J. L. M. Rupp and H. L. Tuller, *Nature Materials*, 2022, **21**, 438–444.
22  G. Walch, B. Rotter, G. C. Brunauer, E. Esmaeili, A. K. Opitz, M. Kubicek, J. Summhammer, K. Ponweiser and J. Fleig, *Journal of Materials Chemistry A*, 2017, **5**, 1637–1649.
23  A. Viernstein, M. Kubicek, M. Morgenbesser, T. M. Huber, E. Ellmeyer, M. Siebenhofer, C. A. F. Vaz and J. Fleig, *Solid State Ionics*, DOI:10.1016/j.ssi.2022.115992.
24  M. Morgenbesser, A. Schmid, A. Viernstein, J. De Dios Sirvent, F. Chiabrera, N. Bodenmüller, S. Taibl, M. Kubicek, F. Baiutti, A. Tarancon and J. Fleig, *Solid State Ionics*, 2021, **368**, 115700.
25  D. Meggiolaro, S. G. Motti, E. Mosconi, A. J. Barker, J. Ball, C. Andrea Riccardo Perini, F. Deschler, A. Petrozza and F. De Angelis, *Energy and Environmental Science*, 2018, **11**, 702–713.
26  J. S. Park, J. Calbo, Y. K. Jung, L. D. Whalley and A. Walsh, *ACS Energy Letters*, 2019, **4**, 1321–1327.
27  C. J. Tong, L. Li, L. M. Liu and O. V. Prezhdo, *Journal of the American Chemical Society*, 2020, **142**, 3060–3068.
28  X. Zhang, M. E. Turiansky, J. X. Shen and C. G. Van De Walle, *Physical Review B*, 2020, **101**, 140101.
29  S. Bitton and N. Tessler, *Energy Environ. Sci.*, 2023, **16**, 2621–2628.
30  D. Moia and J. Maier, *ACS Energy Letters*, 2021, **6**, 1566–1576.
31  T. Kirchartz, J. A. Márquez, M. Stolterfoht and T. Unold, *Advanced Energy Materials*, DOI:10.1002/aenm.201904134.
32  D. Barboni and R. A. De Souza, *Energy and Environmental Science*, 2018, **11**, 3266–3274.
33  S. T. Birkhold, J. T. Precht, R. Giridharagopal, G. E. Eperon, L. Schmidt-mende and D. S. Ginger, *The Journal of Physical Chemistry C*, 2018, **122**, 12633–12639.
34  A. Walsh, D. O. Scanlon, S. Chen, X. G. Gong and S. H. Wei, *Angewandte Chemie - International Edition*, 2015, **54**, 1791–1794.
35  M. Von Smoluchowski, *Z. Phys. Chem.*, 1917, **92**, 129–168.
36  E. Kotomin and V. Kuzovkov, *Reports on Progress in Physics*, 1992, **55**, 2079–2188.
37  A. Zohar, I. Levine, S. Gupta, O. Davidson, D. Azulay and O. Millo, *ACS Energy Letters*, 2017, **2**, 2408–2414.
38  M. T. Neukom, A. Schiller, S. Züfle, E. Knapp, J. Ávila, D. Pérez-del-Rey, C. Dreessen, K. P. S. Zanoni, M. Sessolo, H. J. Bolink and B. Ruhstaller, *ACS Appl. Mater. Interfaces*, 2019, **11**, 23320–23328.

39 W. Clarke, L. J. Bennett, Y. Grudeva, J. M. Foster, G. Richardson and N. E. Courtier, *J Comput Electron*, DOI:10.1007/s10825-022-01988-5.
40 P. Calado, I. Gelmetti, B. Hilton, M. Azzouzi, J. Nelson and P. R. F. Barnes, *J Comput Electron*, 2022, **21**, 960–991.
41 J. Bisquert, *Advanced Energy Materials*, 2024, **2400442**, 1–35.
42 D. Moia, *Phys. Rev. Applied*, 2025, **23**, 014055.

Supporting Information

# Defect chemistry of mixed ionic-electronic conductors under light: halide perovskites as master example

Davide Moia,* Joachim Maier

Max Planck Institute for Solid State Research, Heisenbergstraße 1, 70569 Stuttgart, Germany

*moia.davide@gmail.com
Current address: Fluxim AG, Katharina-Sulzer-Platz 2, 8400 Winterthur, Switzerland

**Table of Contents**

1. Input parameters for calculations

The full list of input parameters used for the calculation shown in Figure 3 in the main text are shown in Table S1. The same parameters are used in all calculations unless stated otherwise

| | | | |
|---|---|---|---|
| $N_C$ | $10^{19} cm^{-3}$ | $K_{\bar{F}}$ | $10^{32} cm^{-6}$ |
| $N_V$ | $10^{19} cm^{-3}$ | $K_{sg,i}$ | $10^{-19} bar^{1/2} cm^3$ |
| $E_g$ | $1.63\ eV$ | $K_{sg,v}$ | $4.38 \times 10^{-5} bar^{1/2} cm^3$ |
| $n_i$ | $2.1 \times 10^5 cm^{-3}$ | $K_{p,i}$ | $10^{-14} cm^3$ |
| $k_{rad}$ | $10^{-11} cm^{-3} s^{-1}$ | $K_{n,v}$ | $10^{14} cm^3$ |
| $\tau_n$ | $2\ \mu s$ | $P(I_2)_{IP}$ | $10^{-2}\ bar$ |
| $\tau_p$ | $2\ \mu s$ | $P(I_2)_{NI}$ | $1.92 \times 10^{-45}$ |
| $n_1 = p_1$ | $2.1 \times 10^5 cm^{-3}$ | $P(I_2)_i$ | $4.38 \times 10^{-24}$ |
| $\gamma_n = \gamma_p$ | $10^{-28} cm^{-6} s^{-1}$ | $\overleftarrow{k}_{\bar{F}}$ | $10^{-15} cm^{-3} s^{-1}$ |
| $E_C - E_{T,i}$ | $0.3\ eV$ | $\vec{k}_{sg,v}$ | $10^{10}\ cm^{-3} s^{-1}$ |
| $E_{T,v} - E_V$ | $0.3\ eV$ | $\vec{k}_{sg,i}$ | $477.8\ bar^{-1/2} s^{-1}$ |

Electronic properties:

- Representative values for bandgap energy and recombination rate constants and lifetimes for MAPI are used (see for example Ref.[1,2])

Ionic properties:

- The value for the anti-Frenkel disorder $K_{\bar{F}} = 10^{32} cm^{-6}$ implies a concentration of defects (vacancies and interstitials) in the intrinsic region of $[I_i'] = [V_I^\cdot] = 10^{16} cm^{-3}$. We note that estimates for the vacancy concentration in MAPI vary in literature, some studies pointing to significantly larger values ($10^{17} - 10^{19}\ cm^{-3}$).[3,4] We show calculation using a lower value to obtain better numerical convergence. The results in the main text and in this document are qualitative relevant also to cases with larger defect concentrations.
- The value of $\overleftarrow{k}_{\bar{F}}$ is estimated based on Equation 9, using a value of $r_{\bar{F}} = 1\ nm$ and $D_{\bar{F}} = 8 \times 10^{-10} cm^{-2} s^{-1}$

Solid-gas exchange:

- The values of $K_{sg,i}$ and $K_{sg,v}$ are selected assuming a trap energy level about 0.3 eV from the band edge for both interstitial (w.r.t. valence band edge) and vacancy defects (w.r.t. conduction band).

The value of $P(I_2)_{IP}$ and $P(I_2)_{NI}$ refer to the iodine partial pressure separating the intrinsic and the P region, and the iodine partial pressure separating the N and the intrinsic region, respectively, on the Kröger-Vink diagram for the equilibrium data. The value of $P(I_2)_{IP}$ is estimated from typical conductivity and mobility values for MAPI. The value of $P(I_2)_{NI}$ is obtained from $P(I_2)_{IP}$, $n_i$ and $K_{\bar{F}}$.

## 2. Shockley-Read-Hall analysis of recombination mediated by mobile ionic defects

We refer to the reactions (n,i), (p,i) in the main text, reported here in Table S1.

Table S1. Redox reactions involving electrons or holes and iodine interstitial defects.

| Redox (interstitial) | $I_i' + h^. \rightleftarrows I_i^\times$ (p, i) | $\vec{R}_{p,i} = \vec{k}_{p,i}[I_i']p$ $\overleftarrow{R}_{p,i} = \overleftarrow{k}_{p,i}[I_i^\times]$ | $K_{p,i} = \frac{[I_i^\times]_{eq}}{[I_i']_{eq} p_{eq}}$ |
|---|---|---|---|
| | $I_i' \rightleftarrows I_i^\times + e'$ (n, i) | $\vec{R}_{n,i} = \vec{k}_{n,i}[I_i']$ $\overleftarrow{R}_{n,i} = \overleftarrow{k}_{n,i}[I_i^\times]n$ | $K_{n,i} = \frac{[I_i^\times]_{eq} n_{eq}}{[I_i']_{eq}} = K_{p,i} K_B$ |

Following the approach in Ref.[5] we can interpret the Shockley-Read-Hall rate based on the rate constants in Table S1. It follows that

$$U_I = \frac{np - n_i^2}{\tau_{n,i}(p + p_{1,i}) + \tau_p(n + n_{1,i})} \quad (S1)$$

where

$$\tau_{n,i} = \left[\overleftarrow{k}_{n,i}\left([I_i^\times] + [I_i']\right)\right]^{-1} = \frac{n_{eq}[I_i^\times]_{eq}}{\vec{k}_{n,i}([I_i^\times] + [I_i'])[I_i']_{eq}} \quad (S2)$$

$$\tau_{p,i} = \left[\vec{k}_{p,i}\left([I_i^\times] + [I_i']\right)\right]^{-1}. \quad (S3)$$

The parameter $\Gamma_{I,i}$ is defined in the main text as the ratio of the forward rate for the (n,i) and (p,i) reactions ($\Gamma_{I,i} = \vec{R}_{p,i}/\vec{R}_{n,i}$) evaluated at $P(I_2) = P(I_2)_i$ at equilibrium.

It follows that $\Gamma_{I,i} = \vec{k}_{p,i} n_i / \vec{k}_{n,i}$. By rearranging the expressions for the capture time constants above, we obtain:

$$\Gamma_{I,i} = \frac{\tau_{n,i}}{\tau_{p,i}} \frac{[I_i']}{[I_i^\times]}\bigg|_i \quad (S4)$$

Where $\frac{[I_i']}{[I_i^\times]}\bigg|_i$ is the ratio of the iodide and iodine interstitial concentrations evaluated at $P(I_2)_i$ and at equilibrium. Based on the Fermi-Dirac statistics that establishes the relative concentration of the negative (occupied by an electron) and neutral defect (unoccupied), we can therefore write

$$\frac{[I_i']}{[I_i^\times]}\bigg|_i = e^{\frac{E_{T,i} - E_i}{k_B T}}, \quad (S5)$$

where the intrinsic energy $E_i = \frac{E_C + E_V}{2} + k_B T \ln\left[\frac{N_V}{N_C}\right]$ and $E_{T,i}$ is the interstitial trap energy associated with the $I_i^\times / I_i'$ redox.

Because $p_{1,i} = N_V e^{\frac{E_V - E_{T,i}}{k_B T}}$ and $n_{1,i} = N_C e^{-\frac{E_C - E_{T,i}}{k_B T}}$, we obtain:

$$\Gamma_{I,i} = \frac{\tau_{n,i}}{\tau_{p,i}} \sqrt{\frac{N_C}{N_V} \frac{n_{1,i}}{p_{1,i}}}. \quad (S6)$$

A similar treatment can be applied to $\Gamma_{I,v}$.

Finally, as mentioned in the main text, the parameters $\Gamma_{p,i}$ and $\Gamma_{n,v}$ are defined to parameterize the rate of hole trapping by an iodide interstitial and electron trapping by an iodide vacancy. Such rates are normalized by the radiative recombination rate, all rates being evaluated at equilibrium and for $P(I_2) = P(I_2)_i$. Their values can therefore be expressed as:

$$\Gamma_{p,i} = \frac{\vec{k}_{p,i}\sqrt{K_{\bar{F}}}}{k_{rad}n_i} \qquad (S7)$$

$$\Gamma_{n,v} = \frac{\vec{k}_{p,v}\sqrt{K_{\bar{F}}}}{k_{rad}n_i}. \qquad (S8)$$

## 3. Discussion of the quasi-equilibrium under light

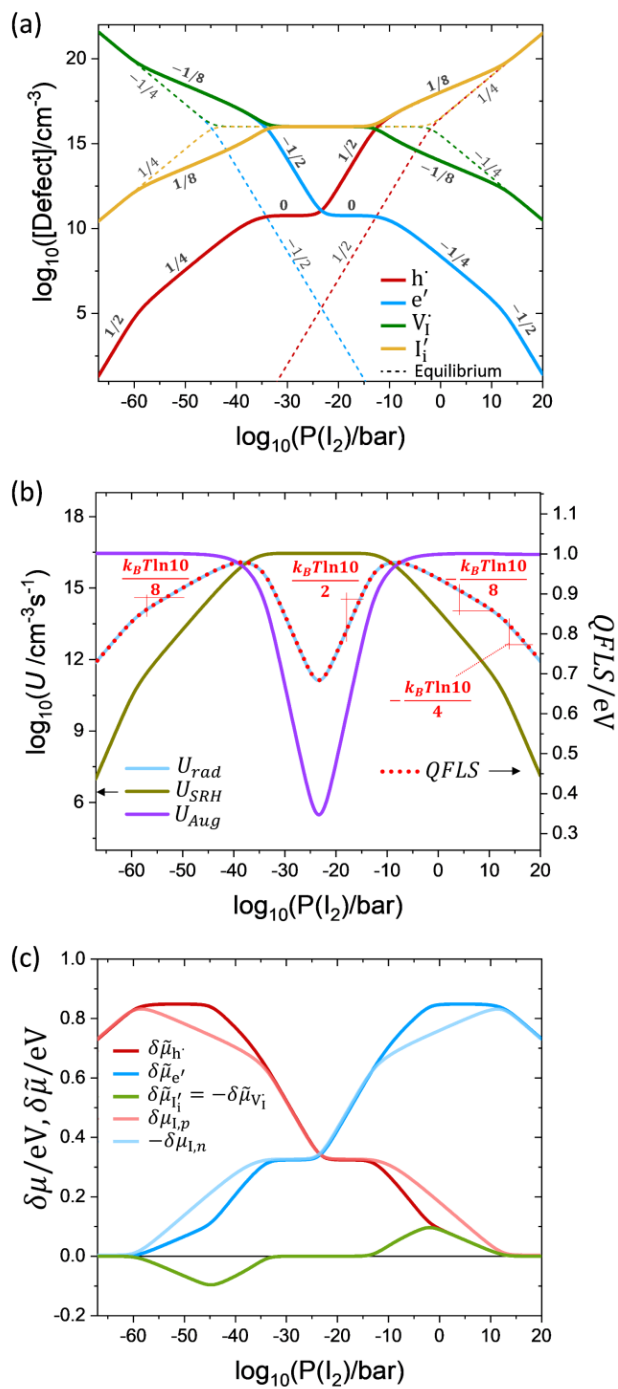

Figure S1. (a) Iodine partial pressure ($P(I_2)$) dependence of the steady-state electronic and ionic defect concentrations in MAPI at equilibrium and under light (~$10^{-5}$ suns) for $\Gamma_{I,i} = 1$ plotted in a Kröger-Vink diagram. (b) Net recombination contributions and quasi-Fermi level splitting. (c) Changes in (quasi-)chemical and (quasi-)electrochemical potentials of all defects and of iodine.

Figure S1a and b show results from a calculation analogous to the one shown in Figure 3, but for a lower light intensity and for a wider $P(I_2)$ range. Figure S1c displays the corresponding changes in chemical and electrochemical potentials of all defects. From the discussion of Figure 3 in the main text, $\delta\tilde{\mu}_{V_I} = -\delta\tilde{\mu}_{I'_i} = -\delta\tilde{\mu}_{I^-}$ ($\Delta\mu_{\bar{F}} = 0$) at all $P(I_2)$, with $\delta\tilde{\mu}_{V_I} \approx 0$ in the intrinsic region under light. In such region, the change in iodine quasi-chemical potentials can be expressed as $\delta\mu_{I,p} \approx \delta\mu_{h^\cdot}$ and $\delta\mu_{I,n} \approx \delta\mu_{e'}$, while in general both electronic and ionic contributions need to be taken into account ($\delta\mu_{I,p} = -\delta\tilde{\mu}_{V_I} + \delta\tilde{\mu}_{h^\cdot}$ and $\delta\mu_{I,n} = -\delta\tilde{\mu}_{V_I} - \delta\tilde{\mu}_{e'}$). Figure S1a shows that the concentration of both iodide defects and of holes (electrons) tend to the equilibrium value for very high (low) $P(I_2)$ values. Consistently, Figure S1c shows that $\delta\mu_{I,p}$ ($\delta\mu_{I,n}$) tends to 0 under such conditions, indicating that holes (electrons) essentially dominate the redox reactions and $\mu^*_{I,p}$ ($\mu^*_{I,n}$) matches $\frac{1}{2}\mu_{I_2,g}$.

Finally, the definition of $\tilde{\mu}^*_{e^-}$ under bias in the main text (see Figure 3c) is meaningful in the context of describing the free enthalpy change in the system per added electron (*i.e.* on increase in $n$ or decrease in $p$). Unlike the quasi-electrochemical potentials $\tilde{\mu}^*_{e'}$ and $\tilde{\mu}^*_{h^\cdot}$, $\tilde{\mu}^*_{e^-}$ has no rigorous meaning when it comes to its relation with the occupation statistics of the electron and hole populations in the material's energy bands. Because anti-Frenkel disorder is at equilibrium in this example even under bias, $\tilde{\mu}^*_{I^-}$ maintains all the properties of $\tilde{\mu}_{I^-,\text{eq}}$.

## 4. One vs two redox-active ionic defects

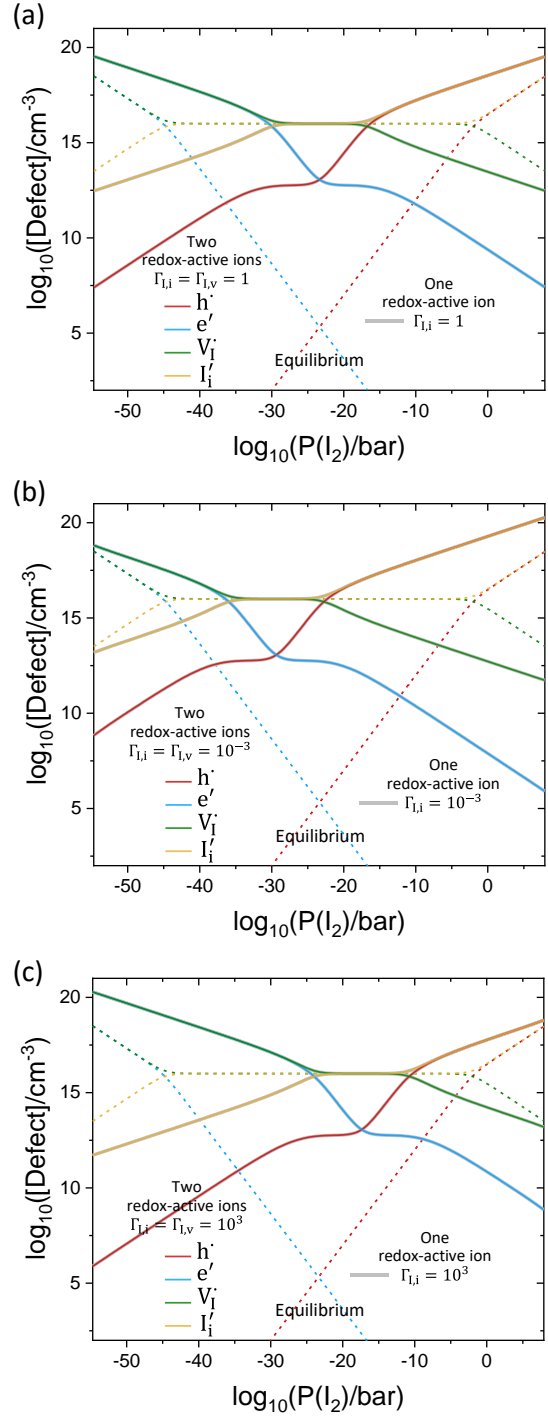

Figure S2. Defect concentrations obtained considering a one- (gray lines) or a two-redox-active mobile ion situation. For the latter, $\Gamma_{I,i} = \Gamma_{I,v}$ is considered. (a) $\Gamma_{I,i} = \Gamma_{I,v} = 1$ (b) $\Gamma_{I,i} = \Gamma_{I,v} = 10^{-3}$, (c) $\Gamma_{I,i} = \Gamma_{I,v} = 10^3$. Illumination of $10^{-3}$ suns is considered. A low value of $\Gamma_{p,i} = \Gamma_{n,v} = 10^{-2}$ is used.

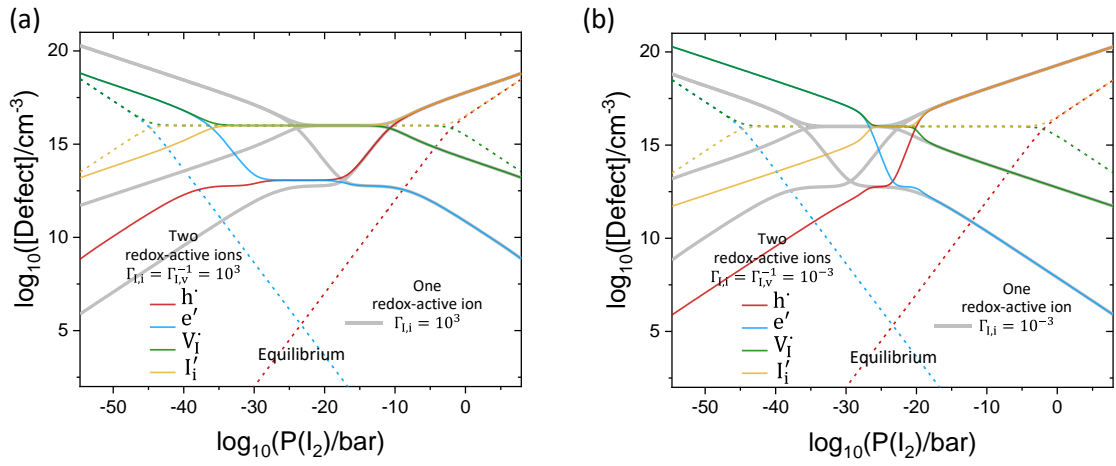

Figure S3. Defect concentrations obtained considering a one- (gray lines) or a two-redox-active mobile ion situation. The influence of $\Gamma_{I,i} \neq \Gamma_{I,v}$ is shown for (a) $\Gamma_{I,i} = 10^3, \Gamma_{I,v} = 10^{-3}$ and (b) $\Gamma_{I,i} = 10^{-3}, \Gamma_{I,v} = 10^3$. Illumination of $10^{-3}$ suns is considered.

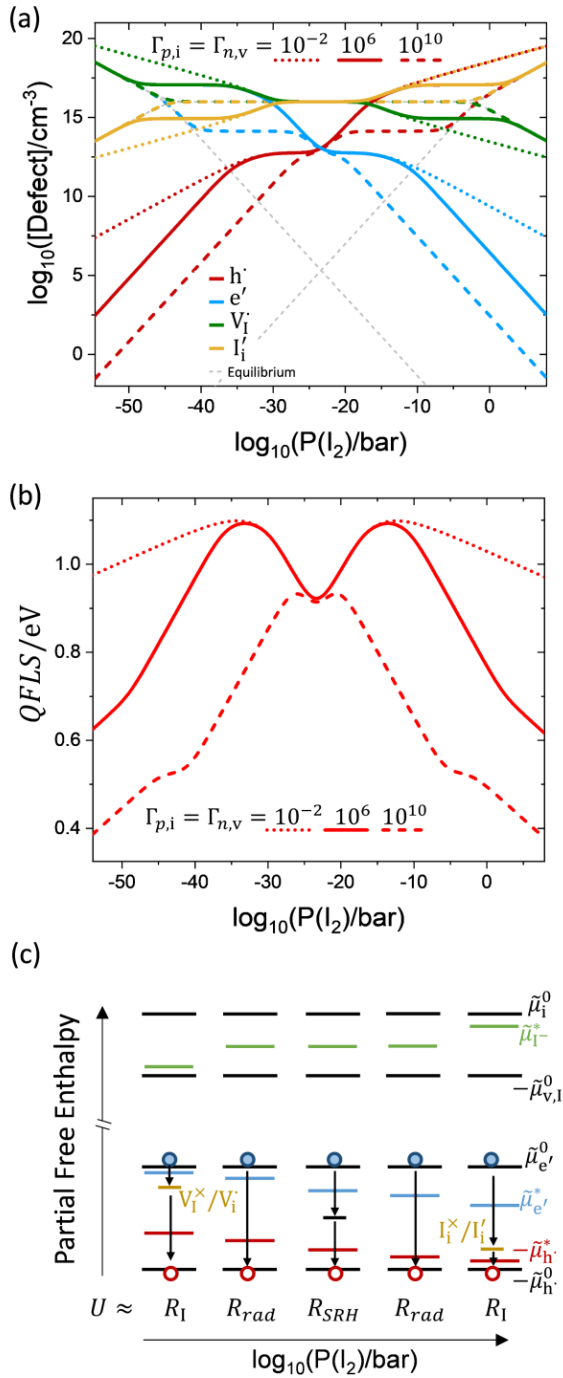

Figure S4. (a) Defect concentration and (b) $QFLS$ for a MAPI film calculated for $\Gamma_{p,i} = \Gamma_{n,v} = 10^{-2}, 10^6$ and $10^{10}$ (corresponding to $\vec{k}_{p,i} = 2.1 \times 10^{-24}, 2.1 \times 10^{-16}$ and $2.1 \times 10^{-12} \ cm^3 s^{-1}$). Illumination of $10^{-3}$ suns and $\Gamma_{I,i} = \Gamma_{I,v} = 1$ are considered. (c) Generalized energy diagram showing the dominant recombination mechanisms for different $P(I_2)$ regions in (a) and (b) ($\Gamma_{p,i} = \Gamma_{n,v} = 10^6$ case). The $I_i^\times/I_i'$ and the $V_I^\times/V_I^\cdot$ energy levels are included (~0.3 eV from the valence band maximum and from the conduction band minimum, respectively).

Figure S4 shows that, when large values of $\Gamma_{p,i}$ and $\Gamma_{n,v}$ are used, the presence of the additional recombination pathway provided by the second defect leads to a similar trend as the one observed for Figure 4, but where a drop in $QFLS$ is observed not only for the high (recombination via $I_i^\times/I_i'$) but also for the low $P(I_2)$ range (recombination via $V_I^\times/V_I^\cdot$). This is illustrated schematically in Figure S4c.

## 5. Effect of solid-gas exchange kinetics on the ionic and electronic quasi-equilibrium

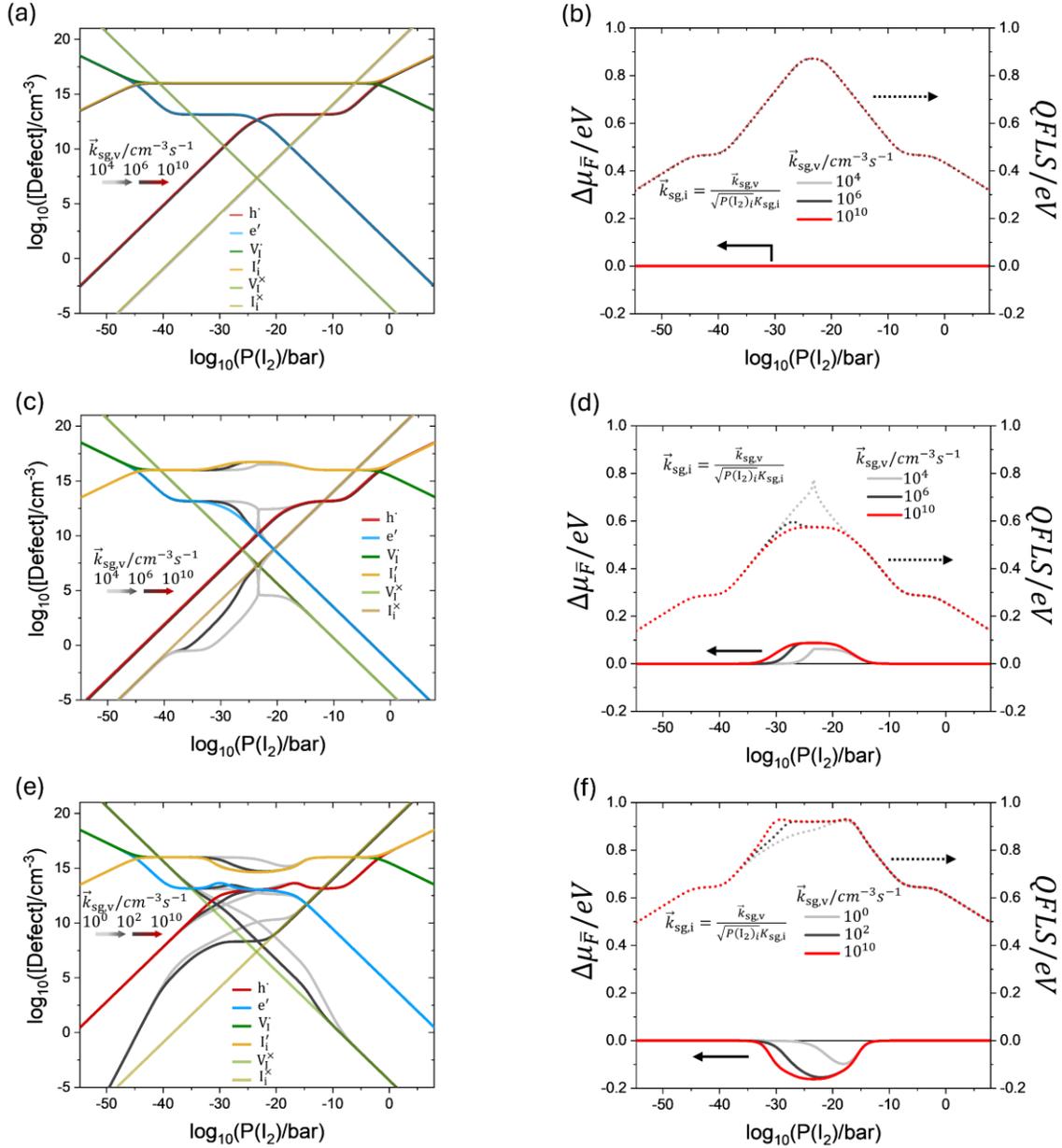

Figure S5. (left column) Defect concentrations and (right column) ionic ($\Delta\mu_{\bar{F}}$) and electronic chemical potentials ($QFLS$) for different values of the rate constants controlling the solid-gas exchange. (a, b) $\Gamma_{I,i} = \Gamma_{I,v} = 1$, (c, d) $\Gamma_{I,i} = 10^{-3}, \Gamma_{I,v} = 10^{3}$, (e, f) $\Gamma_{I,i} = 10^{3}, \Gamma_{I,v} = 10^{-3}$. Illumination of $10^{-3}$ suns is considered. Deviation from the *sg-eq* behavior trend for the neutral ionic defect concentration highlights the kinetic limitations of the solid-gas exchange.

## 6. Effect of redox reaction kinetics on the ionic and electronic quasi-equilibrium

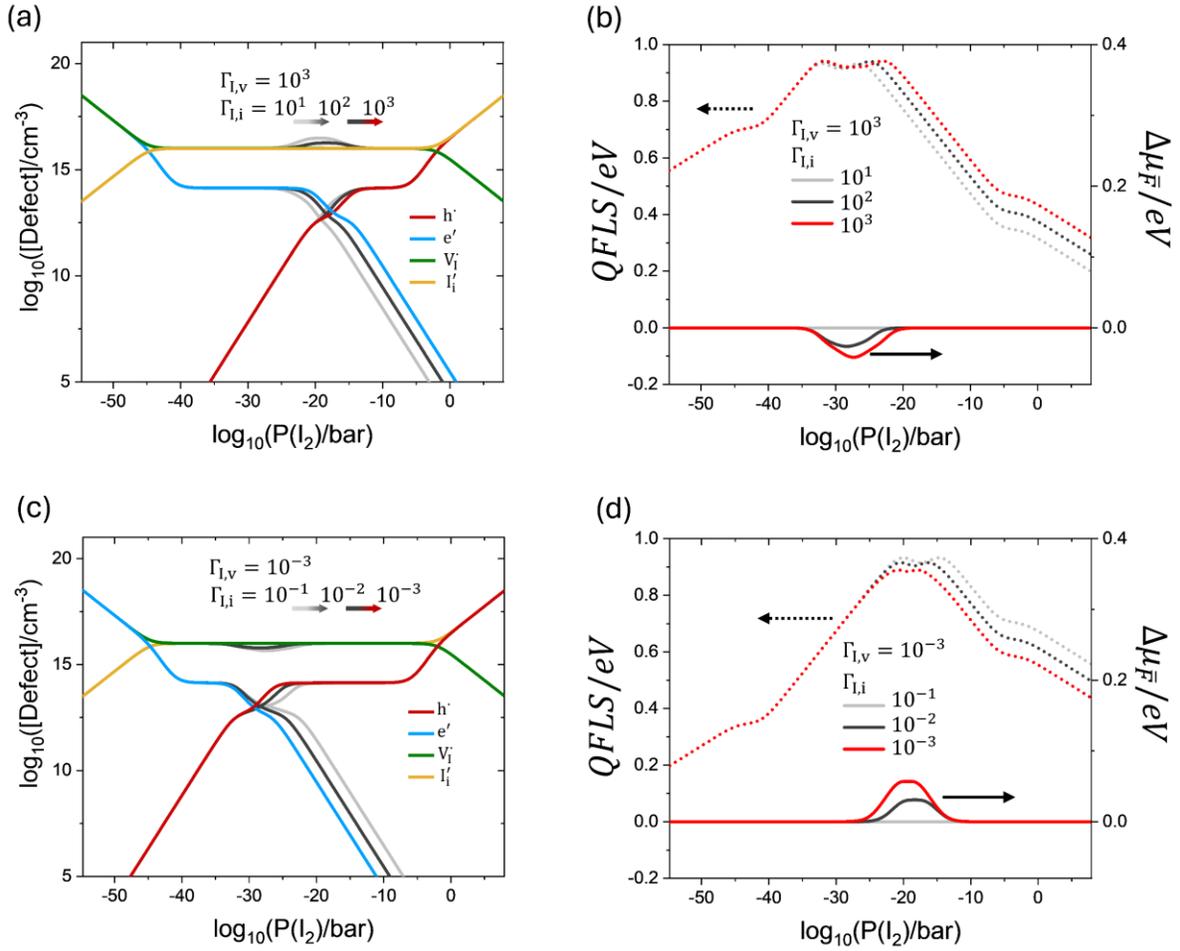

Figure S6. (left column) Defect concentrations and (right column) ionic ($\Delta\mu_{\bar{F}}$) and electronic chemical potentials ($QFLS$) for: (a, b) $\Gamma_{I,v} = 10^3$ (c, d) $\Gamma_{I,v} = 10^{-3}$ and varying values of $\Gamma_{I,i}$. Illumination of $10^{-3}$ suns is considered.

## 7. Electronic and ionic quasi-equilibrium in absence of immobile defect-mediated recombination

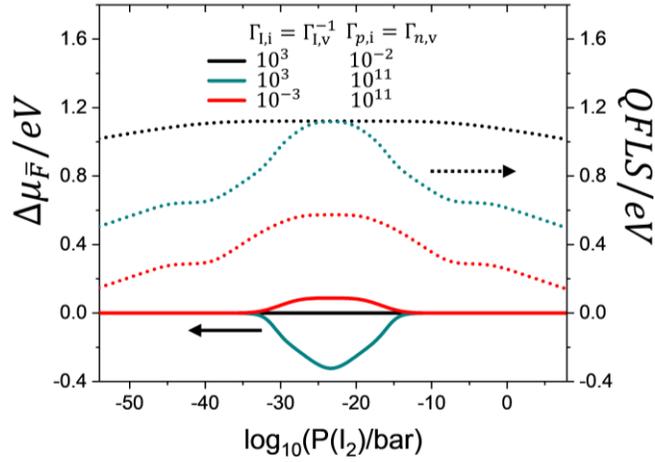

Figure S7. ionic ($\Delta\mu_{\bar{F}}$) and electronic chemical potentials ($QFLS$) for different input parameters. Illumination of $10^{-3}$ suns is considered. As no immobile trap-mediated SRH recombination is considered, when recombination via iodide defects is significant (large $\Gamma_{p,i} = \Gamma_{n,v}$), the $QFLS$ profile correlates with the density of anti-Frenkel defects. The latter scale exponentially with $\Delta\mu_{\bar{F}}$.

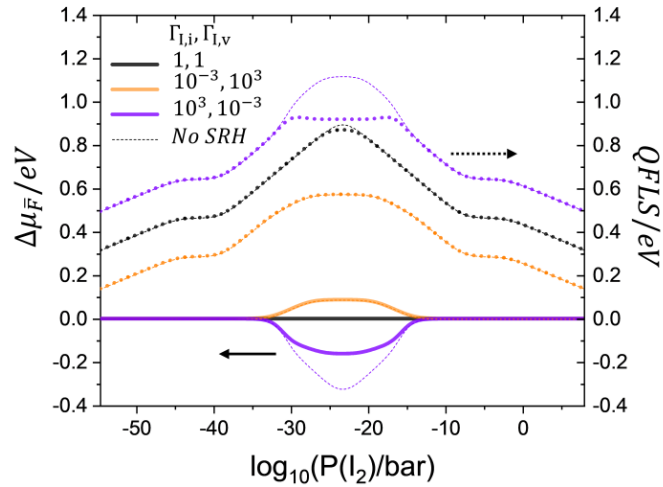

Figure S8. Data for the ionic ($\Delta\mu_{\bar{F}}$) and electronic chemical potentials ($QFLS$) shown in Figure 6a in the main text, compared with the analogous situation but without contribution from SRH recombination due to immobile defects.

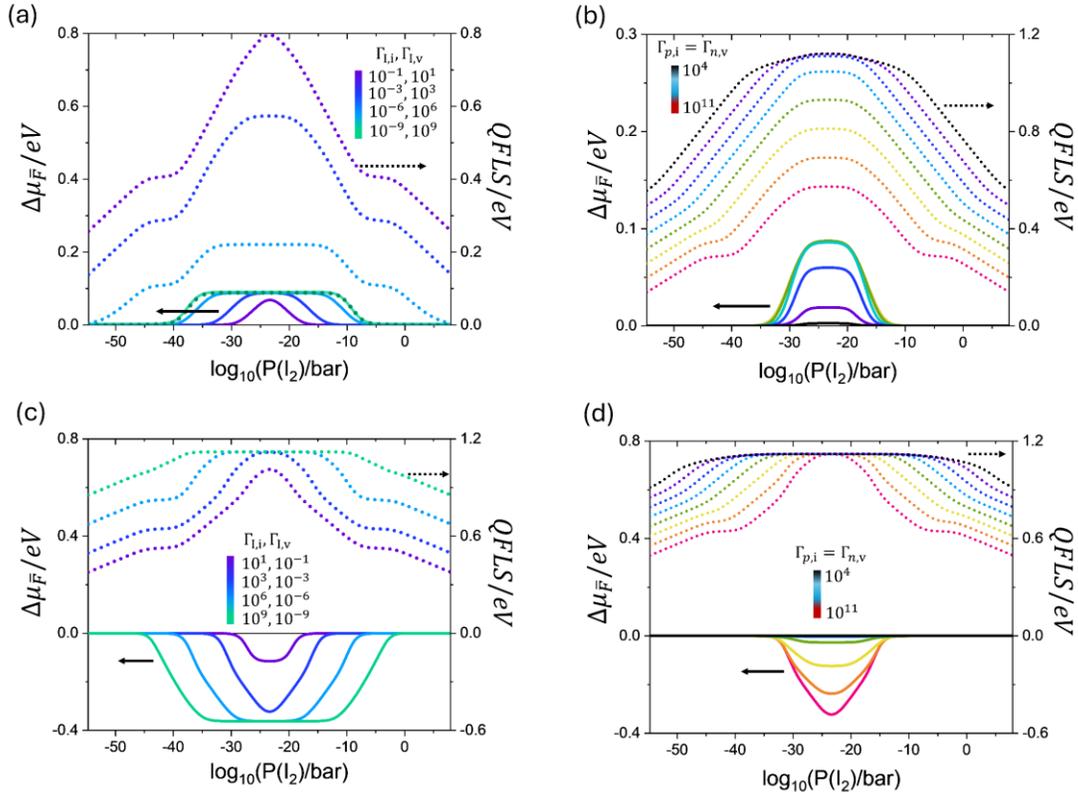

Figure S9. Dependence of the ionic ($\Delta\mu_{\bar{F}}$) and electronic ($QFLS$) change in chemical potential on the $\Gamma_{I,i}$, $\Gamma_{I,v}$, and $\Gamma_{p,i}$, $\Gamma_{n,v}$ parameters, as a function of P(I$_2$). Trends for (a) increasing $h_{ion} > 0$ ($\Gamma_{p,i} = \Gamma_{n,v} = 10^{11}$), (b) increasing $\Gamma_{p,i} = \Gamma_{n,v}$ ($\Gamma_{I,i} = 10^{-3}$, $\Gamma_{I,v} = 10^{3}$), (c) decreasing $h_{ion} < 0$ ($\Gamma_{p,i} = \Gamma_{n,v} = 10^{11}$), (d) increasing $\Gamma_{p,i} = \Gamma_{n,v}$ ($\Gamma_{I,i} = 10^{3}$, $\Gamma_{I,v} = 10^{-3}$). All remaining parameters are the same as the ones used in Figure 3. $h_{ion} = \log_{10}\left(\frac{\Gamma_{I,v}}{\Gamma_{I,i}}\right)$. The data are obtained for a situation analogous to Figure 8, but with no contribution from SRH recombination due to immobile defects.

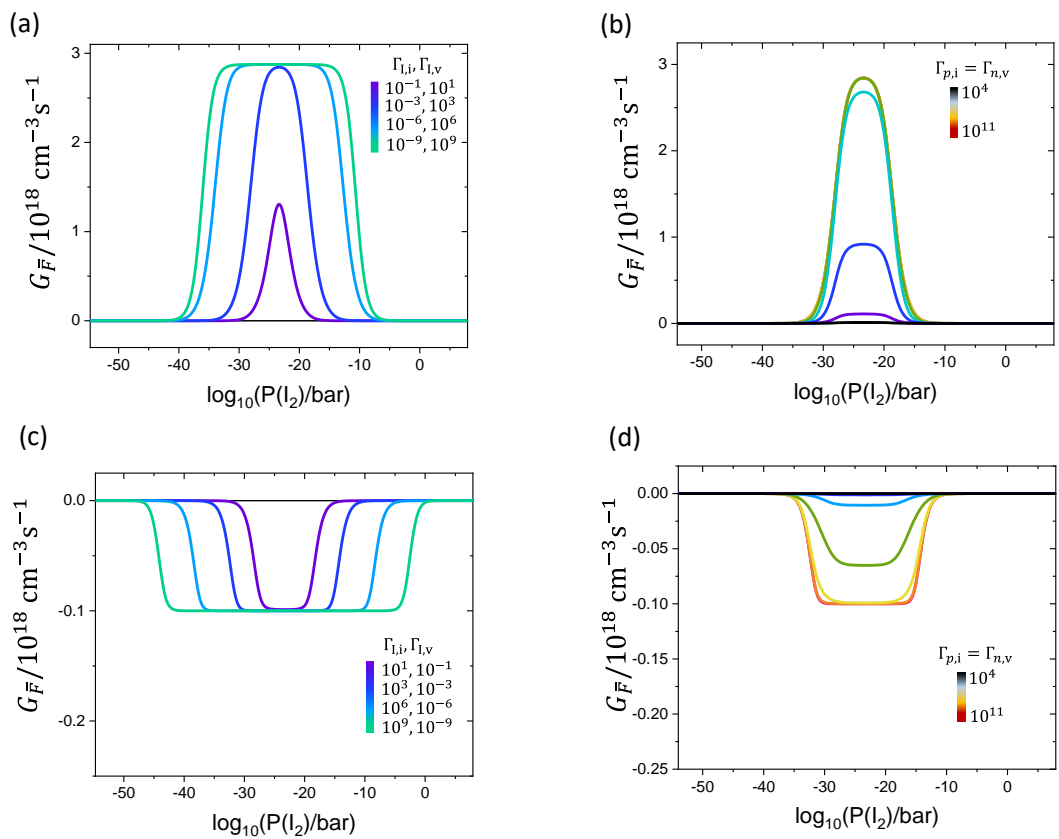

Figure S10. (a), (b), (c) and (d) show the effective ionic generation rates for the data displayed in Figure S9a, b, c and d, respectively.

## 8. Supporting References

1 T. Kirchartz, J. A. Márquez, M. Stolterfoht and T. Unold, *Advanced Energy Materials*, 2020, **10**, 1904134.
2 M. E. Ziffer, J. C. Mohammed and D. S. Ginger, *ACS Photonics*, 2016, **3**, 1060–1068.
3 D. A. Jacobs, C. M. Wolff, X.-Y. Chin, K. Artuk, C. Ballif and Q. Jeangros, *Energy Environ. Sci.*, 2022, **15**, 5324–5339.
4 D. Moia, M. Jung, Y.-R. Wang and J. Maier, *Phys. Chem. Chem. Phys.*, 2023, **25**, 13335–13350.
5 J. Maier, *Physical Chemistry of Ionic Materials*, Wiley-VCH Verlag GmbH & Co. KGaA, 2nd edn., 2023.